\documentclass[twocolumn]{aa} 

\usepackage{graphicx}
\usepackage{txfonts}
\usepackage{rotating}
\usepackage{aas_macros}
\usepackage{threeparttable}
\usepackage[shortcuts]{extdash}
\usepackage{float}
\usepackage{bm}
\usepackage{mathtools,xparse}
\usepackage[utf8]{inputenc}
\usepackage[T1]{fontenc}
\usepackage{pdfpages}
\usepackage{lipsum}
\usepackage{adjustbox}
\usepackage{amsmath}
\usepackage{natbib}
\newcommand{\msun}{\,\mathrm{M}_\odot}

\DeclareUnicodeCharacter{00A0}{ }

\DeclarePairedDelimiter{\norm}{\lVert}{\rVert}
\NewDocumentCommand{\normL}{ s O{} m }{%
  \IfBooleanTF{#1}{\norm*{#3}}{\norm[#2]{#3}}_{L_2($\Omega$)}%
}

\begin{document}

\title{Asteroid
mass estimation with the robust adaptive Metropolis algorithm}
   \author{L. Siltala
          \inst{1,2}\thanks{Corresponding author, e-mail: lauri.siltala@helsinki.fi}
          \and
          M. Granvik\inst{1,3}}

   \institute{
   Department of Physics, P.O. Box 64, FI-00014 University of Helsinki, Finland
         \and
            Nordic Optical Telescope, Apartado 474, E-38700 S/C de La Palma, Santa Cruz de Tenerife, Spain
         \and
         Division of Space Technology, Luleå University of Technology, Kiruna, Box 848, S-98128, Sweden}
    


\abstract{
The bulk density of an asteroid informs us about its interior structure and composition. To constrain the bulk density one needs an estimate for the mass of the asteroid. The mass is estimated by analyzing an asteroid's gravitational interaction with another object, such as another asteroid during a close encounter. An estimate for the mass has typically been obtained with linearized least-squares methods despite the fact that this family of methods is not able to properly describe non-Gaussian parameter distributions. In addition, the uncertainties reported for asteroid masses in the literature are sometimes inconsistent with each other and suspected to be unrealistically low.}
{We present a Markov-chain Monte Carlo (MCMC) algorithm for the asteroid mass estimation problem based on asteroid-asteroid close encounters. We verify that our algorithm works correctly by applying it to synthetic data sets. We then use astrometry available through the Minor Planet Center to estimate masses for a select few example cases and compare our results to results reported in the literature.}
{Our mass estimation method is based on the robust adaptive Metropolis algorithm that has been implemented into the OpenOrb asteroid orbit computation software. Our method has the built-in capability to analyze multiple perturbing asteroids and test asteroids simultaneously.}
{We find that our mass estimates for the synthetic data sets are fully consistent with the ground truth. The nominal masses for real example cases typically agree with the literature but tend to have greater uncertainties than what is reported in  recent literature. Possible reasons for this include different astrometric datasets and/or weights, different test asteroids, different force models and/or different algorithms. For (16) Psyche, the target of NASA's Psyche mission, our maximum likelihood mass is approximately 55\% of what is reported in the literature. Such a low mass would imply that the bulk density is significantly lower than previously expected and hence disagrees with the theory of (16) Psyche being the metallic core of a protoplanet. We do, however, note that masses reported in recent literature remain within our 3-sigma limits.}
{The new MCMC mass-estimation algorithm performs as expected, but a rigorous comparison with results from a least-squares algorithm with the exact same dataset remains to be done. The matters of uncertainties in comparison with other algorithms and correlations of observations also warrant further investigation.}

\maketitle

\section{Introduction}

We describe, validate, and apply a novel Markov-chain Monte Carlo (MCMC) method for the estimation of asteroid masses based on close encounters between two or more asteroids. These new advances build on our previous work on the subject \citep{Sil17}, which showed that MCMC methods do not share the potential weaknesses of the other methods used for estimating asteroid masses such as misleading estimates for the random component of the total uncertainty budget.

An estimate for an asteroid's mass is typically obtained by analyzing how it perturbs the orbits of other so-called test bodies that it has
close encounters with. In practice, the test body is either another
asteroid or a spacecraft. Asteroid mass estimation based on planetary ephemerides is another possibility. Finally, in the case of binary asteroids, mass estimation may also be carried out by examining the orbit of the secondary asteroid around the primary. A comprehensive review of these approaches is provided by \citet{Car12}.

Asteroid mass estimation based on close encounters between asteroids has traditionally been done with least-squares schemes. Applying the least-squares method implies that one has to make certain assumptions regarding the shape of the probability distributions of the model parameters to describe the model parameters' uncertainties. This is problematic as these assumptions, e.g., that the uncertainties can be described by Gaussian distributions, have not been validated. In addition, uncertainties of the mass are often underestimated in the literature \citep{Car12}.

Recent work by other authors using the asteroid-asteroid close encounter approach has focused on including multiple asteroids (both massive perturbers and massless test asteroids) simultaneously in the computations (e.g., \citet{Bae17} and \citet{Gof14}, the latter of which simultaneously used an impressive total number of 349 737 simultaneous asteroids) as opposed to the more traditional approach of computing separate masses for different test asteroids where possible, and taking the average of these as the final perturber mass.

Recent developments in the area also include work on correcting astrometric biases from star catalogues \citep{Far15}, thus improving the quality of astrometric observations. The recent second Gaia data release \citep{DR2} includes astrometry of unprecedented accuracy for a large number of Solar System objects \citep{DR2_Fed}. In addition, the DR2 star catalogue allows for re-reducing Earth-based data, again to minimize biases resulting from star catalogues, thus leading to potential improvements in older data. In particular, the use of Gaia data promises significant improvements in asteroid masses in the near future.

Finally, there are ongoing efforts to improve asteroid shape models and thus volumes \citep{Vii18,Ver18}. When both the mass and the volume are known, the bulk density can be trivially calculated.

Our new mass estimation method, which is implemented using the OpenOrb asteroid-orbit-computation software \citep{Gra09}, features 
\begin{itemize}
\item a proper observational weighting scheme \citep{Bae17}, 
\item star catalogue debiasing corrections \citep{Far15},
\item the option to account for gravitational perturbations by other asteroids \citep{Bae17}, 
\item an MCMC algorithm based on the robust adaptive Metropolis \citep{Vih12} and the global adaptive scaling with adaptive Metropolis \citep{And08} algorithms,
\item an improved outlier detection algorithm compared to \citet{Sil17}, and
\item the option to use multiple test asteroids and/or perturbing asteroids instead of just one of each as in \citet{Sil17}.
\end{itemize}

In what follows we first describe our method. We then validate our method with synthetic astrometry and, finally, compare our results for a few real cases with the results found in the literature including our own previous results.

\section{Theory and methods}

\subsection{Problem statement}

The case of asteroid-asteroid perturbations leads to a
multi-dimensional inverse problem where the aim is to solve for the masses
of the perturbing asteroids as well as six orbital elements at a specific
epoch for both the perturbing asteroids and the test asteroids
by fitting the model predictions to astrometric observations taken over a relatively long
timespan. We parameterize the
orbits with heliocentric Cartesian state vectors at a
specific epoch, that is, $\bm{S} =
(x,y,z,\dot{x},\dot{y},\dot{z})$ where $(x,y,z)$ is the asteroid's
position and $(\dot{x},\dot{y},\dot{z})$ its velocity at epoch $t_0$. In
general, the total set of model parameters is thus
$\bm{P}=(\bm{S}_1,\bm{S}_2,\ldots,\bm{S}_{N_\mathrm{obj}},M_1,M_2,\ldots,M_{N_\mathrm{per}})$,
where $M_i$ is the mass of the $i$th perturber, $N_\mathrm{obj}$ is the number of asteroids included, and $N_\mathrm{per}$ is
the number of perturbers considered. Hence, $N_\mathrm{obj} \ge N_\mathrm{per}$.

We use the $\chi^2$ test statistic in matrix notation to represent the goodness of fit
 resulting from a set of model parameters $\bm{P}$:
\begin{equation}\label{chi2_definition}
\chi^2 = \sum_{i=1}^{N_\mathrm{obj}} \sum_{j=1}^{N_{\mathrm{obs},i}} \left[\bm{\epsilon}^T_{i,j}\bm{\Sigma}^{-1}_{i,j}\bm{\epsilon}_{i,j}\right]
\end{equation}
where $\bm{\epsilon_{i,j}}$ is a column vector consisting of $O - C$ residuals:
\begin{equation}
\bm{\epsilon}_{i,j}({\alpha},{\delta}) = \left((\alpha_{i,j}^0 -
  \alpha_{i,j}(\bm{P}))\cos{\delta_{i,j}^0},\;\delta_{i,j}^0 -
  \delta_{i,j}(\bm{P})\right)
 \end{equation}
 
$N_{\mathrm{obs},i}$ is the number of observations used for asteroid
$i$, $\alpha_{i,j}^0$ and $\delta_{i,j}^0$ are the observed Right
Ascension (RA) and Declination (Dec), respectively, of asteroid $i$
at time $t_j$, $\alpha_{i,j}(\bm{P})$ and $\delta_{i,j}(\bm{P})$ are
the predicted RA and Dec of asteroid $i$ at time $t_j$ and $\bm{\Sigma}_{i,j}^{-1}$ is the information matrix, that is, the inverse of the covariance matrix, of a given observation.
To compute $\alpha_{i,j}(\bm{P})$ and $\delta_{i,j}(\bm{P})$ we
integrate the orbits of the asteroids through the observational
timespan while taking the gravitational perturbations of both the
perturbing asteroids and the planets into account. The closer the predicted
positions are to the observations, the smaller is the $\chi^2$ statistic. Smaller $\chi^2$ statistics thus correspond to a better agreement
between observations and model prediction.


\subsection{Markov-chain Monte Carlo}
The general idea of an MCMC algorithm is
to create a Markov chain to estimate the unknown posterior probability
distributions of the parameters $p(\bm{P})$ of a given model. A Markov
chain is a construct consisting of a series of elements in
which each element is derived from the one preceding it. In a properly
constructed Markov chain the posterior distributions of individual
elements in the chain match the probability distributions of these
elements. Thus as the end result of MCMC, one gets the probability
distributions of each parameter in the model. From these
distributions, one can directly determine the maximum-likelihood
values from the peaks of the distributions alongside the credible intervals. The main advantage of MCMC methods is the rigorous mapping from the observational uncertainty to the uncertainty in orbital elements and mass. As mentioned in the Introduction, it is
common to assume a symmetric, Gaussian shape for the probability distributions, but as we showed in our previous paper \citep{Sil17}, the probability distribution of the mass is in some cases non-symmetric and such a distribution cannot be correctly described if assuming a Gaussian shape.

The MCMC method used here consists of two separate phases. For the first 5000 accepted solutions
we use global adaptive scaling with adaptive Metropolis \citep[GASWAM;][Algorithm 4]{And08}, which combines
the earlier adaptive Metropolis \citep[AM;][]{Haa02} with the idea of also adapting the scaling parameter that multiplies the covariance matrix of the proposal distribution
in an attempt to coerce the acceptance rate to a desired percentage. In the second phase we switch to
the robust adaptive Metropolis (RAM) algorithm
\citep{Vih12}, which constantly adapts the shapes of the proposal distributions, rather than just the scaling factors, to achieve a desired proposal acceptance rate. The reason for the algorithm change during the run is
that, in our experience, the RAM algorithm occasionally has problems with suboptimal initial values that result in an extended burn-in phase and poor mixing. GASWAM, on the other hand, is able to better deal with the suboptimal initial values and is used to produce improved initial values and proposal distributions for RAM.

The proposed parameters $\bm{P}'$ are generated by adding
deviates $\bm{\Delta P}$ to the previously accepted, or $i$th, set of
parameters $\bm{P}_i$:
\begin{equation}
 \bm{P}' = \bm{P}_i + \bm{\Delta P}\,.
\end{equation}
The deviates are computed as
\begin{equation}
 \bm{\Delta P} = \bm{S_i} \bm{R}\,,
\end{equation}
where $\bm{S_i}$ is the Cholesky decomposition of the proposal
distribution and $\bm{R}$ is a
$(6N_\mathrm{obj}+N_\mathrm{per})$-vector consisting
of Gaussian-distributed random numbers. At this point,
proposals with negative masses are automatically rejected as they
are not physically plausible while all masses greater than or equal
to zero are permitted.

Once a physically-plausible proposal has been generated, we integrate the corresponding orbits through the observational timespan
and calculate the $\chi^2$ statistics (Eq.~\ref{chi2_definition}) corresponding to the proposal. The posterior probability density $p$ is then obtained as
\begin{equation}\label{eq:posterior}
  p(\bm{P}') \propto \exp(-\frac{1}{2} \chi^2(\bm{P}')) \,.
\end{equation}
Next, the posterior probability density is compared to the previously
accepted solution:
\begin{equation}\label{eq:ratio}
  a_i = \frac{p(\bm{P}')}{p(\bm{P}_i)} = \exp(-\frac{1}{2} \left(\chi^2(\bm{P}') - \chi^2(\bm{P})\right))
\end{equation}
We use a non-informative constant prior distribution for both the orbital elements (in Cartesian elements) and the masses (in solar masses) in Eq.~\ref{eq:posterior}. For a matter of simplicity we use a value of unity, but one can choose an arbitrary value because the values will cancel each other out in Eq.~\ref{eq:ratio}. The choice of the prior distribution will not have a significant impact on the results when the target asteroids have extensive and/or accurate astrometry available, and the posterior distributions are therefore relatively well constrained. The choice of the prior distribution becomes more important when there is not enough astrometry available to properly constrain the posterior distributions with the likelihood function alone \citep[cf.][and Solin, Granvik, Farnocchia,~in preparation]{FARNOCCHIA201518}. In practice the choice of the prior distribution mostly affects the posterior distribution for the perturber mass rather than the orbital elements of the perturber(s) or the test asteroid(s), because an accurate modeling of the close encounter requires that the orbits have to be known to a high accuracy for the mass-estimation approach to make sense. In what follows we will primarily focus on example cases for which the prior distribution can safely be assumed to have a negligible effect and leave a detailed discussion of the choice of the prior distribution for future work.

If $a_i > 1$, the proposed solution is better than the previously
accepted solution and hence it is automatically accepted as the next
transition. Otherwise, it is accepted with a probability of $a_i$. The
first proposal in the chain is always accepted, because there is no
previous solution to compare with.

The proposal distribution $\bm{S_i}$ is constantly updated after each proposal, whether
it is accepted or not, based on the computed chain so far. For the first phase with less than 5,000 accepted
transitions we use the GASWAM formula \citep{And08}. For the proposal distribution we derive an empirical
covariance matrix from the Markov chain and then scale it by a factor $\lambda$ \citep[cf.][]{Sil17}:
  \begin{equation}
    \bm{S}_i\bm{S}_i^T = \lambda_n \frac{1}{i - 1} \sum_{j=1}^i (\bm{P_j} -
    \overline{\bm{P}})(\bm{P_j} - \overline{\bm{P}})^{\rm{T}} + \epsilon
    \bm{I_d}\,,
\end{equation}
Here $\bm{P}_j$ represents all of the accepted solutions in the chain so far,
$\overline{\bm{P}}$ represents their mean, $\bm{I_d}$ is the identity
matrix, and $\epsilon$ is an arbitrary small parameter. We empirically
found that $\epsilon=10^{-26}$ produces good results and that the
results are not particularly sensitive to its value. Even $\epsilon=0$ can work in practice, but the ergodicity has only been mathematically proven for $\epsilon>0$ \citep{Haa02}.
For the initial value of $\bm{S}_n\bm{S}_n^T$ we use the covariance matrices for each asteroid orbit and mass considered combined into
a single block matrix.

In GASWAM the scaling parameter $\lambda_n$ is no longer constant as in AM, but adapted
constantly. We calculate $\lambda_n$ as follows \citep{And08}:
\begin{equation}
\lambda_n = \lambda_{n-1}+n^{-0.5}*(a_n-a_*)
\end{equation}
Here $n$ represents all proposals so far, including those not accepted. $a_n$ represents the acceptance probability
of the last proposal as described earlier while $a_*$ represents the desired mean acceptance rate, for which we use the
standard value of $0.234$ \citep{Rob97}. We also note that due to the nature of these equations, the covariance matrix itself
is only updated when a new proposal is accepted while its scaling parameter is updated after every proposal.

During the first phase of the chain we also apply our outlier rejection algorithm. At points $i = 50, 500, 5000$ 
we reject observations with Mahalanobis distances greater than 4. Mahalanobis distance is defined as 
\begin{equation}
D_M(\alpha,\delta) = \sqrt{\bm{\epsilon}_{i,j}({\alpha},{\delta})
\bm{S}^{-1}_{i,j}\bm{\epsilon}_{i,j}({\alpha},{\delta})}
\end{equation}

where $\bm{\epsilon}_{i,j}({\alpha},{\delta})$ represents the mean accepted residuals for a specific observation while $\bm{S}^{-1}_{i,j}$ represents the inverse of the covariance matrix of this observation.
We force acceptance of the immediately following proposal
to prevent the chain from getting stuck which may otherwise happen due to the outlier rejection having a significant impact 
on $\chi^2$. Mathematically this is equivalent to starting a new chain and it should have little significance 
in the actual results, as we remove the burn-in phase from the results. 

For the second phase of the chain where $i > 5000$, we switch to the RAM update formula \citep{Vih12} instead, using
the final $\bm{S}_i$ from the first phase as our initial proposal distribution:
\begin{equation}
   \bm{S}_n\bm{S}_n^T = \bm{S}_{n-1}\left(\bm{I} + \eta_n(a_n -a_*){\bm{\Delta P}_n\bm{\Delta P}_n^T \over \norm{\bm{\Delta P}_n}^2}\right)\bm{S}_{n-1}^T
\end{equation}
where $n$ represents the total amount of proposals so far, $\bm{I}$ represents the identity matrix, $\eta_n$ is a step size sequence that can be arbitrarily
chosen, for which we selected $n^{-0.5}$ as suggested in the original paper of \citet{Vih12}, $a_n$ is the
acceptance probability of the last proposal as described above, $a_*$ is the desired mean acceptance probability
for which we again use the standard value of $0.234$, 
and, finally, $\bm{\Delta P}_n$ represents the proposal deviates as described above.

We repeat the process until the desired number of transitions, typically 25,000, is
reached. A new chain is then started with the initial masses of
$2\bm{M}_\mathrm{init}$ and the same orbital elements as used to initiate
the first chain. This is done both to ensure that a sufficiently large
range of masses is tested, and to ensure that the parameters converge
to the same posterior distribution with different starting values.

We determine our credible intervals by calculating a
  kernel-density estimate (KDE) based on the statistics of repetitions
  such that the limits encompassing 68.26\% of the probability mass 
  around the peak of the KDE correspond to $1\sigma$ while the limits
  encompassing 99.73\% of the probability mass correspond to
  $3\sigma$.

  \subsection{Initial values}

We obtain the initial orbits and covariance matrices using the least-squares method separately for each asteroid. Our initial values for the masses of the perturbing asteroids are computed from their $H$ magnitudes assuming a spherical shape \citep{Che04}:
\begin{equation}
 \bm{M}_\mathrm{init} = \frac{\pi}{6}\rho \bm{D}^{\circ 3}\,,
\end{equation}
where we assume that the bulk density $\rho = 2.5$g/cm$^3$ and ${}^{\circ 3}$ represents the Hadamard power. The diameter $\bm{D}$ is \citep{Che04}:
\begin{equation}
 \bm{D} = 1329 \times 10^{-\bm{H}/5}\,p_V^{-1/2}\,\mathrm{km}\,,
\end{equation}
where we assume that the geometric albedo $p_V = 0.15$.

 As a consequence of the initial masses not having been computed with the least-squares method, we do not have covariance information for the masses. For the initial covariance matrix we assume that the correlations between the mass and the orbital elements are zero while the variance of the mass is assumed to be $10^{-12} \times \bm{M}_\mathrm{init}$, which we have empirically tested to be sufficient for our cases.
The use of suboptimal initial values for the variances and covariances including the masses is not a significant issue, because the adaptive algorithms will rapidly update the matrix with improved values.

\section{Data}

We test our mass estimation algorithm with both synthetic and real astrometry.  Synthetic astrometry is useful for verifying the mathematical consistency and functionality of the algorithm, because we know the exact masses of the perturbers as well as the noise that was applied to the astrometry. On the other hand, real astrometry will provide physically meaningful results that can be compared to the literature. The real data also allows us to gauge the performance of our algorithm in practice. We describe data weighting scheme used in this work as well as both data sets in the following subsections.

\subsection{Weighting of the astrometry}

In our previous work we used root-mean-square (RMS) values of the residuals for each
individual asteroid as the weights of each observation for the given asteroid in our model, which is hardly
a realistic approach given that data points from different observatories have different uncertainties.
The issue has largely been corrected as we
now use the observational error (that is, standard deviations for the right ascension and declination) developed by \citet{Bae17}, itself an updated version of the earlier \citet{Bae11b} model, for the
astrometry from all observatories considered in the model. 

For data from observatories not included in the Baer model we assume the
following uncertainties depending on the observation date $t_j$:
\begin{equation}
 \sigma_{\alpha,i,j} = \sigma_{\delta,i,j} =
 \begin{cases}
  3.0 & t_j < 1890        \\
  2.0 & 1890 < t_j < 1950 \\
  1.5 & 1950 < t_j < 1990 \\
  1.0 & 1990 < t_j < 2010 \\
  0.6 & t_j > 2010
 \end{cases}
\end{equation}
Here the pre-1990 weights equal the suggested weights for photographic
observations of \citet{Far15} while the post-1990 weights are chosen
by us based on experience. A significant benefit of using an observational
error model is that while in our previous paper we only used
astrometry obtained after 1990-01-01.0, because older photographic
observations likely have significantly larger errors than modern CCD
observations, we can now study cases with greater
observational timespans than before and this is expected to lead to reduced uncertainties in the resulting asteroid masses.

We multiply our weights from the error model for $N$ observations of a single
asteroid by a single observatory on a given night by $\sqrt{N}$ so as
to take observational correlations into account (see e.g.
\citet{Far15}) 
and debias the observations with the model of \citet{Far15} for each observation for which the debiasing corrections are available. We note that the Baer model also includes correlations for some observatories, but in their place we have opted for the $\sqrt{N}$ factor as it can be used universally for all observatories
unlike the Baer model.



\subsection{Synthetic astrometry}

We generated synthetic astrometry based on seven separate test cases detailed in Table~\ref{synth_results}. We first obtained astrometry from the Minor Planet Center, then applied the above-mentioned observational weighting model to the data, 
and finally calculated orbits for each asteroid using the least-squares method in the same manner as with the real data. 
Using these initial orbits we then computed ephemerides for each object corresponding to the epoch of each real observation 
for said object. The total number of observations and the observation sequence of the synthetic observations are thus identical to the real observations 
for these objects. To generate the ephemerides we account for the perturbing asteroid(s) involved in each case and use the nominal masses reported by \citet{Car12} for each perturbing asteroid. We chose not to include perturbations 
by the planets or other perturbing asteroids so as to not slow down the mass estimation.
Finally, we add random Gaussian noise to each computed position using standard deviations corresponding to our observational 
weights for each individual observation.


\subsection{Real astrometry}

We chose several different encounters between asteroids for this work. First, we chose several test cases based on those on our previous paper \citep{Sil17} by selecting most of the asteroids we previously studied and added a second massless test asteroid for each studied perturber. For these targets we obtained all of the astrometry available through the Minor Planet Center during the observational time span as described in Table~\ref{astpairs_1990}. We will refer to this data set as the real data set.

\begin{table*}
\footnotesize
  \caption{The list of mutual encounters between asteroids used in this work. Our notation is such that 
   the asteroid(s) on the left side of the semicolon are considered perturbers while 
   those on the right side are massless test asteroids. The reference values are weighted averages of previous literature values taken from the work of \protect \citet{Car12}. The third column contains references that describe the encounters. We note that \citet{Gal02} opted not to publish some of these encounters in the referred paper, and such encounters were obtained from the authors' website directly (http://staryweb.fmph.uniba.sk/index.php?id=2171) The fourth and fifth column respectively denote the start and end date of the observational time span for the perturbing asteroid. For the test asteroids we use all of the available astrometry within this timespan. References. (1) \citet{Bae17}; (2) \citet{Gal02}; (3) Zielenbach, private communication.}
  \label{astpairs_1990}
  \begin{center}
    \begin{tabular}{ccccc}
      \hline
      Encounter & Ref.~mass.~1  & Reference & $t_0$ (MJD) & $t_n$ (MJD)\\
      &                                                          [$10^{-11} \msun$] \\ 
      \hline
          {[\textbf{7},88;17799,52443,7629,11701]} & $0.649 \pm 1.06$ & (1) & 47911 & 58154\\ 
          {[10;1259,57493]}   &  $4.34 \pm 0.26$ & (1) & 48023 & 58083 \\ 
          {[15;765,14401]}    & $1.58 \pm 0.09$ & (2) & 48138 & 58317 \\ 
          {[16;17799,20837]}   & $1.37 \pm 0.38$ & (2) & 47894 & 58337\\ 
          {[16;91495,151878]}  & $1.37 \pm 0.38$ & (1) & 47894 & 58337 \\ 
          {[19;3486,27799]}    & $0.433 \pm 0.073$ & (1) & 48087 & 58172  \\ 
          {[29;987,7060]}      & $0.649 \pm 0.101$ & (1) & 47985 & 57450 \\ 
          {[41;8212,10332]}    & $0.317  \pm 0.06$ & (3) & 48072 & 58374 \\ 
          {[52;124,306]}       & $1.20 \pm 0.29$ & (1) & 47204 & 57982 \\
          {[7,\textbf{88};17799,52443,7629,11701]} & $0.769 \pm 0.156$& (1)  & 47911 & 58154 \\
          {[89;38057,54846]}   & $0.337  \pm 0.092$ & (3) & 47907 & 58158.  \\ 
          {[216;23747,170964]}  & $0.233  \pm 0.10$ & (3) & 47894 & 58130\\ 
          {[704;43993,48500]}   & $1.65  \pm 0.23$ & (1) & 48020 & 58493\\ 

      \hline
    \end{tabular}
  \end{center}
\end{table*}

\section{Results and discussion}

\subsection{Synthetic data}

The masses that synthetic data was generated with typically fall within the $1\sigma$ boundaries of our results and the probability distributions are more or less symmetric (Table~\ref{synth_results}). Our maximum-likelihood (ML) masses are not exactly the same as the true masses, but are in most cases extremely 
close and some differences are expected due to the noise added to the synthetic data. Asteroids (52) Europa and (704) Interamnia have relatively wide uncertainties in comparison to the other test cases and the explanation is that the chosen encounters for these asteroids provide relatively weak constraints on the masses. This also explains the relatively large difference between the ML mass and the true mass for (704) Interamnia, although even in this case the true mass is within the $3\sigma$ boundaries. The non-Gaussian mass distribution for (52) Europa may simply be due to the algorithm only permitting positive masses---the lower 3$\sigma$ limit of (52) Europa can in practice be considered to be zero as the design of our algorithm makes it impossible to reach values of exactly zero.

\begin{table*}
\footnotesize
  \caption{Compilation of the MCMC algorithm's results for all synthetic
    encounters. The correct masses are those the synthetic data sets were generated with, thus we know them to be 
    exact. In the case with two perturbers (7 and 88) the perturber masses are on separate rows with each row corresponding 
    to the perturber emphasized with bold text.
    The third column is the value of the cumulative distribution function of the mass corresponding to the correct mass.}
  \label{synth_results}
  \begin{center}
    \begin{tabular}{cccccc}
      \hline
      Encounter	   & Correct mass & CDF & ML mass              & $1\sigma$ boundaries   & $3\sigma$ boundaries   \\
      & [$10^{-11} \msun$] & & [$10^{-11} \msun$] & [$10^{-11} \msun$]     & [$10^{-11} \msun$]  \\ \hline       
      {[\textbf{7},88;17186,46529,52443,7629,11701,142429]} & 0.649  & 0.54 & 0.647 & [0.608, 0.685] & [0.531, 0.759]   \\
      {[10;3946,6006,11215,24433]}                          & 4.34  & 0.94  & 4.23  & [4.12, 4.33]   & [3.93, 4.53]    \\
      {[15;765,3591,14401,24066]}                           & 1.58  & 0.85  & 1.44  & [1.29, 1.59]   & [0.995, 1.88]    \\
      {[16;6442,13206,19462,20837]}                         & 1.37	& 0.40  & 1.46	& [1.10, 1.80]   & [0.411, 2.50]    \\
          {[19;3486,27799,113990,114525]}                   & 0.433 & 0.23	& 0.491	& [0.405, 0.580] & [0.233, 0.755]    \\
      {[52;124,8269,19790,22336]}                           & 1.20  & 0.41	& 1.29  & [0.488, 2.11]  & [0.00104, 4.58]  	\\
      {[7,\textbf{88};17186,46529,52443,7629,11701,142429]} & 0.769 & 0.89  & 0.678 & [0.594, 0.758] & [0.424, 0.918]  \\
      {[704;1467,7461,10034,48500]}                         & 1.65  & 0.02	& 3.47	& [2.52, 4.43]   & [0.593, 6.30] 	  \\  \hline
    \end{tabular}
  \end{center}
\end{table*}

\begin{figure}
  \begin{center}
    \includegraphics[width=1.0\columnwidth]{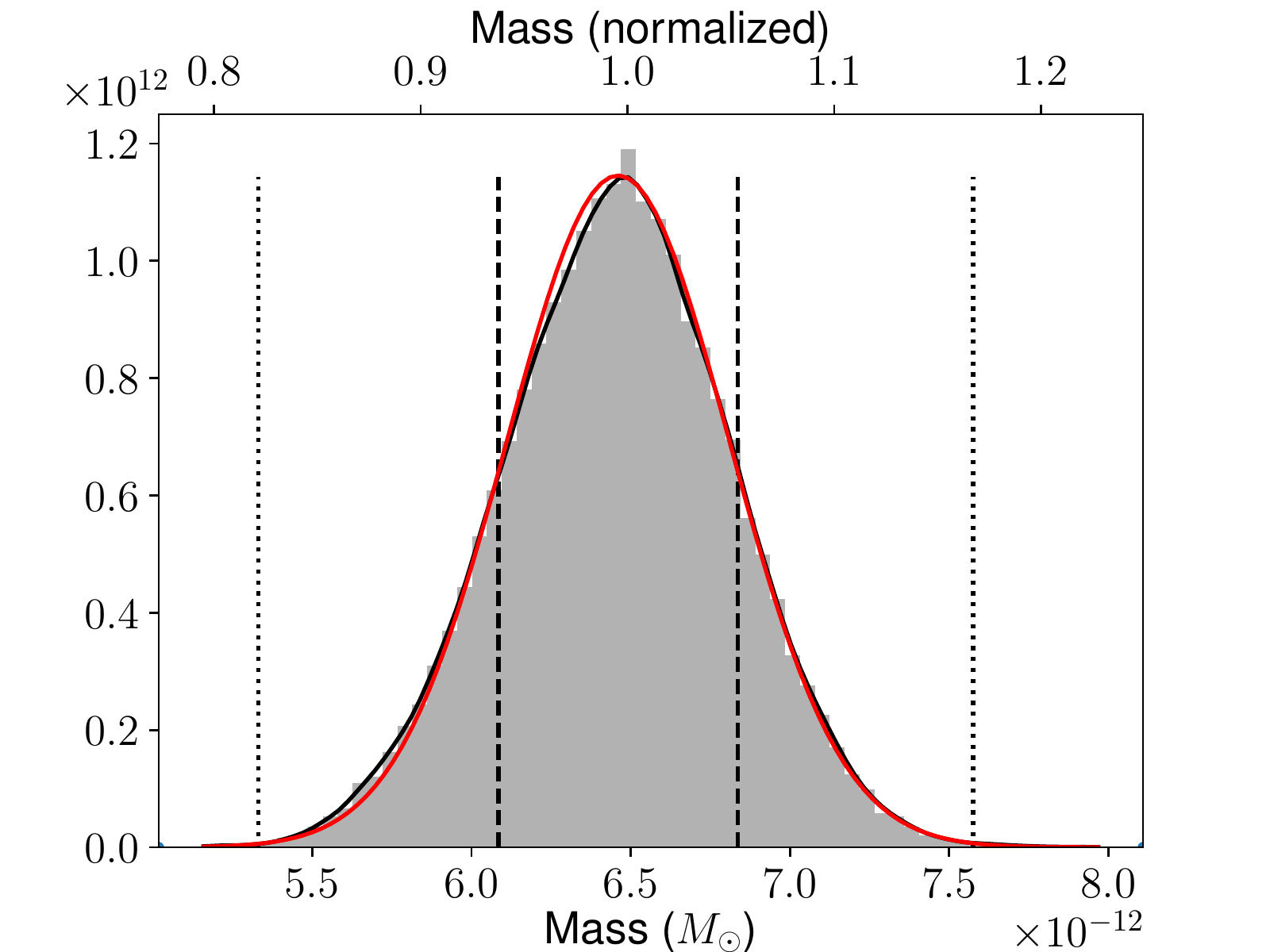}
    \caption{Results of the MCMC algorithm 
      applied to the synthetic [\textbf{7},88;17186,46529,52443,7629,11701,142429] encounter for asteroid (7) Iris. The upper x-axis is
      normalized such that 1.0 equals the exact mass the data was generated with. The black dashed and dotted 
      vertical lines represent the $1\sigma$ and $3\sigma$ limits respectively while the red graph corresponds to a Gaussian distribution based on the median and standard deviation of the results.}
    \label{7_synth}
  \end{center}
\end{figure}

\begin{figure}
  \begin{center}
    \includegraphics[width=1.0\columnwidth]{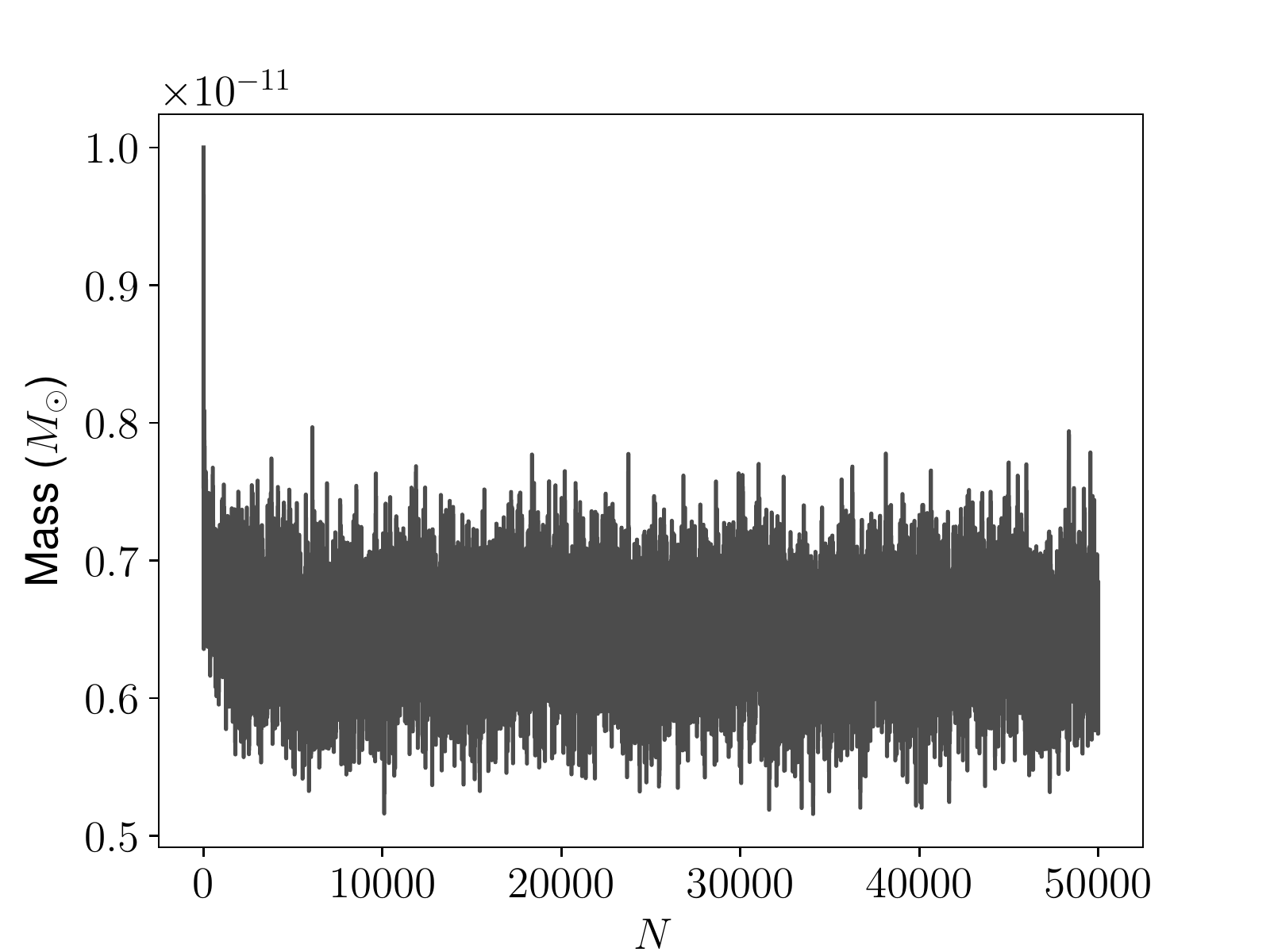}
    \caption{Trace of the MCMC chain of the synthetic [\textbf{7},88;17186,46529,52443,7629,11701,142429] encounter in terms of mass of asteroid (7) Iris.}
    \label{7_synth_trace}
  \end{center}
\end{figure}

The probability distribution for the mass of (7) Iris serves as an example of an excellent, practically Gaussian result in terms of the similarity between the true mass and the estimated mass (Fig.~\ref{7_synth}). The case of [7,88;17186,46529,52443,7629,11701,142429] is special in that it has two separate perturbing asteroids
due to the perturbers themselves having comparable masses and several close encounters with each other during the observational timespan. The mixing of the Markov chain is fairly good 
and also the power of adaptive methods is particularly clear in Figure \ref{7_synth_trace}: 
in the very beginning of the run, the proposal distribution for mass is quite
narrow with only very small jumps between accepted proposals, and as the chain progresses, the distribution grows wider and the jumps 
larger due to our adaptation scheme before quickly converging to an optimal distribution .

To further investigate the impact that additional perturbing and/or test asteroids have on the results, we investigated the mass of (7) Iris based on synthetic data with different combinations of asteroids. In the synthetic case,
a single test asteroid is already sufficient to obtain very good mass estimates for (7) Iris, while the inclusion of additional test asteroids reduces the uncertainty of the results (Table~\ref{iris_test}). In addition, it is apparent that in this case the mass estimates for both (7) Iris and (88) Thisbe can be obtained based on their mutual perturbations alone, and no massless test asteroids are strictly necessary. The uncertainties resulting are, however, relatively wide suggesting that their mutual perturbations are weak in comparison to those with their test asteroids. Finally, we see that it is not actually necessary to consider the perturbations of (88) Thisbe on (7) Iris to get good mass estimates of the latter. Nonetheless, including (88) Thisbe does further reduce the uncertainties of the mass of (7) Iris.

\begin{table*}
\footnotesize
  \caption{MCMC results for (7) Iris and (88) Thisbe based on synthetic data with different combinations of asteroids.}
  \label{iris_test}
  \begin{center}
    \begin{tabular}{ccccc}
      \hline
      Encounter	  & Correct mass & ML mass              & $1\sigma$ boundaries   & $3\sigma$ boundaries    \\
      & [$10^{-11} \msun$] & [$10^{-11} \msun$]   & [$10^{-11} \msun$]   & [$10^{-11} \msun$]  \\ \hline       
      {[\textbf{7},88;17186,46529,52443,7629,11701,142429]} & 0.649  & 0.647 & [0.608, 0.685] & [0.531,  0.759]   \\
      {[7,\textbf{88};17186,46529,52443,7629,11701,142429]} & 0.749  & 0.678 & [0.594, 0.758] & [0.424,  0.918]   \\
      {[\textbf{7},88]}                                    & 0.649  & 0.444 & [0.103, 0.936] & [0.000606, 2.43]  \\
      {[7,\textbf{88}]}                                    & 0.749  & 0.615 & [0.387, 0.848] & [0.000809, 1.36]  \\
      {[7;17186]}                                          & 0.649  & 0.640 & [0.567, 0.714] & [0.427, 0.862]    \\
      {[7;17186,46529]}                                   &  0.649  & 0.633 & [0.575, 0.689] & [0.456, 0.802]    \\
      {[7;17186,46529,52443]}                            & 0.649  & 0.649 & [0.597, 0.704] & [0.489, 0.805]   
    \end{tabular}
  \end{center}
\end{table*}

\subsection{Real data}

For our real data set we did two separate MCMC runs for each case, one including the perturbations of Ceres, Pallas, and Vesta via the BC430 ephemerides \citep{Bae17} (Table~\ref{main_results_bc430}) and one without (Table~\ref{main_results_nobc430}). These three asteroids were selected as they are by far the three most massive asteroids. In theory, the BC430 ephemeris permits us to include a significantly larger number of perturbers with pre-determined masses at the cost of increased computation time, but we expect that doing so will yield diminishing returns due to the remaining asteroids having significantly lower masses. This is something we intend to investigate in greater detail in future work. For this study we have checked the literature to ensure that our target asteroids have no published close encounters with other perturbers during the observation epoch.
The run without BC430 perturbations was done to estimate the impact that asteroid perturbers can have on the results. The results including the BC430 perturbations should be considered our final results, because those results are based on a more accurate dynamical model. As with the previous synthetic test cases, a very small lower $3\sigma$ boundary can in practice be considered to be zero.

\begin{sidewaystable}
\begin{threeparttable}
\footnotesize
  \caption{Compilation of the MCMC algorithm's results including the perturbations of Ceres, Pallas, and Vesta for our real data set.
    The first reference masses and their uncertainties are from \citet{Sil17} whereas the second are weighted averages of previous literature values \citep{Car12}. In the case of (7) Iris and (88) Thisbe, the result refers to the designation in bold text.}
  \label{main_results_bc430} 
  \begin{center}
  \renewcommand{\arraystretch}{1.2}
    \begin{tabular}{cccccccccc} 
      Encounter	  & ML mass              & $1\sigma$ boundaries   & $3\sigma$ boundaries  & Ref. mass~1  & $1\sigma$ boundaries   & $3\sigma$ boundaries & Ref. mass~2  & Diameter & Density \\
      & [$10^{-11} \msun$] & [$10^{-11} \msun$]   & [$10^{-11} \msun$]   & [$10^{-11} \msun$] & [$10^{-11} \msun$] & [$10^{-11} \msun$] & [$10^{-11} \msun$ & [km] & [g/cm$^3$] \\ \hline 

          {[\textbf{7},88;17186,52443,7629,11701]}     & 0.316  & [0.194, 0.437]   & [0.00259, 0.801]&      0.210	 & [0.0499, 0.354] & [0.000660, 1.20]   & $0.649 \pm 1.06$ & $216 \pm 7$ (1) & $1.19^{+0.44}_{-0.47}$ \\ 
          {[7;17186,52443]}     & 0.623  & [0.463, 0.778]   & [0.129, 1.09]&      0.210	 & [0.0499, 0.354] & [0.000660, 1.20]   & $0.649 \pm 1.06$  & $216 \pm 7$ (1) & $2.35^{+0.63}_{-0.64}$  \\
          {[10;1259,57493]}   & 5.43  & [3.69, 7.08]   & [0.281, 10.04]    & 2.48 & [2.21, 2.77] & [1.63, 3.32]    & $4.34 \pm 0.26$  & $411 \pm 20$ (2) & $2.97^{+1.00}_{-1.05}$ \\

          {[15;765,14401]}     & 1.40  & [1.20, 1.62]   & [0.792, 2.03]&      1.11	 & [0.914, 1.25]  & [0.574, 1.61]   & $1.58 \pm 0.09$  & $275 \pm 5$ (1) & $2.56^{+0.43}_{-0.39}$ \\ 
          {[16;17799,20837]}  & 0.783  & [0.434, 1.13]   & [0.00580, 2.25]&       n/a	 & n/a  & n/a   & $1.37 \pm 0.38$  & $223 \pm 7$ (3)& $2.68^{+1.21}_{-1.22}$ \\ 
        {[16;91495;151878]}  & 0.743  & [0.466, 1.02]   & [0.00910, 1.62]&       n/a	 & n/a  & n/a   & $1.37 \pm 0.38$  & $223 \pm 7$ (3)  & $2.54^{+0.98}_{-0.98}$ \\
          {[19;3486,27799]}   & 0.391 & [0.195, 0.602] & [0.000345, 1.20] & 0.141 & [0.0567, 0.285]  & [0.000467, 0.828] & $0.433 \pm 0.073$   & $211 \pm 4$ (2)  &  $1.58^{+0.86}_{-0.80}$ \\ 
          {[29;987,7060]}     & 0.541 & [0.00694, 1.42] & [0.00151, 5.79]  & 0.258	& [0.0163, 0.898]  & [0.00238, 4.43] & $0.649 \pm 0.101$  & $204 \pm 3$ (2) & $2.42^{+3.93}_{-2.39}$ \\
          {[41;8212,10332]}   & 0.527 & [0.0633, 1.32] & [0.00168, 3.69]  & n/a & n/a  & n/a  & $0.317  \pm 0.06$  & $188 \pm 5$ (2) & $3.01^{+4.54}_{-2.66}$ \\ 
          {[52;124,306]}      & 1.24  & [0.00385, 3.29]  & [0.00384, 10.6] & 0.893 & [0.232, 1.91]        & [0.00319, 6.05]  & $1.20 \pm 0.29$   & $314 \pm 5$ (2) & $1.52^{+2.51}_{-1.52}$ \\ 
           {[7,\textbf{88};17186,52443,7629,11701]}     & 0.606  & [0.285, 0.97]   & [0.000492, 1.76]&      n/a	 & n/a & n/a   & $0.769 \pm 0.156$  & $212 \pm 10$ (2) & $2.42^{+1.49}_{-1.32}$\\ 
          {[88;7629,11701]}  & 0.745  & [0.151, 1.74]   & [0.00149, 5.9]&       n/a	 & n/a  & n/a   & $0.769 \pm 0.156$  & $212 \pm 10$ (2) & $2.97^{+3.98}_{-2.40}$ \\ 
          {[89;38057,54846]}   & 0.921 & [0.261, 1.85] & [0.00139, 5.01]  & n/a & n/a  & n/a  & $0.337  \pm 0.092$  & $142 \pm 4$ (2) & $12.22^{12.36}_{-8.81}$ \\ 
          {[216;23747,170964]}   & 0.0283 & [0.00121, 0.0927] & [0.00121, 0.471]  & n/a & n/a  & n/a  & $0.233  \pm 0.10$   & $121 \pm 5$ (2) & $0.61^{+1.38}_{-0.59}$ \\ 
         {[704;43993,48500]}   & 1.46 & [0.65, 2.33] & [0.00240, 4.97]  & 0.155 & [0.00910, 0.664]  & [0.00182, 3.50]  & $1.65  \pm 0.23$  & $317 \pm 5$ (4) & $4.49^{+1.58}_{-1.59}$ \\ 
    \end{tabular}
    \begin{tablenotes}
    \item 1) \citep{Vii17}; 2) \citep{Han17}; 3) \citep{Dru18}; 4) \citep{Car12}

   \end{tablenotes}
  \end{center}
\end{threeparttable}
\end{sidewaystable}

\begin{sidewaystable}
\footnotesize
  \caption{Compilation of the MCMC algorithm's results without the perturbations of Ceres, Pallas, and Vesta for our real data set.
    The first reference masses and their uncertainties are from \citet{Sil17} whereas the second are weighted averages of previous literature values \citep{Car12}}
  \label{main_results_nobc430} 
  \begin{center}
    \begin{tabular}{cccccccc}
      \hline
      Encounter	  & ML mass              & $1\sigma$ boundaries   & $3\sigma$ boundaries  & Ref. mass~1  & $1\sigma$ boundaries   & $3\sigma$ boundaries & Ref. mass~2  \\
      & [$10^{-11} \msun$] & [$10^{-11} \msun$]   & [$10^{-11} \msun$]   & [$10^{-11} \msun$] & [$10^{-11} \msun$] & [$10^{-11} \msun$] & [$10^{-11} \msun$] \\ \hline 

          {[\textbf{7},88;17186,52443,7629,11701]}     & 0.478  & [0.343, 0.605]   & [0.0689, 0.851]&      0.210	 & [0.0499, 0.354] & [0.000660, 1.20]   & $0.649 \pm 1.06$ \\
          
          {[7;17186,52443]}     & 0.510  & [0.352, 0.669]   & [0.0481, 1.01]&      0.210	 & [0.0499, 0.354] & [0.000660, 1.20]   & $0.649 \pm 1.06$ \\
          {[10;1259,57493]}   & 7.01  & [5.27, 8.80]   & [1.54, 12.1]    & 2.48 & [2.21, 2.77] & [1.63, 3.32]    & $4.34 \pm 0.26$ \\ 

          {[15;765,14401]}     & 1.55  & [1.33, 1.77]   & [0.91, 2.21]&      1.11	 & [0.914, 1.25]  & [0.574, 1.61]   & $1.58 \pm 0.09$ \\ 
          {[16;17799,20837]}  & 0.851  & [0.492, 1.21]   & [0.000775, 2.06]&       n/a	 & n/a  & n/a   & $1.37 \pm 0.38$ \\ 
          {[19;3486,27799]}   & 0.262 & [0.0993, 0.454] & [0.000311, 1.04] & 0.141 & [0.0567, 0.285]  & [0.000467, 0.828] & $0.433 \pm 0.073$   \\ 
          {[29;987,7060]}     & 0.533 & [0.00128, 1.41] & [0.00128, 5.11]  & 0.258	& [0.0163, 0.898]  & [0.00238, 4.43] & $0.649 \pm 0.101$  \\ 
          {[41;8212,10332]}   & 1.05 & [0.302, 2.07] & [0.00136, 5.25]  & n/a & n/a  & n/a  & $0.317  \pm 0.06$  \\ 
          {[52;124,306]}      & 1.29  & [0.0295, 3.42]  & [0.00342, 13.4] & 0.893 & [0.232, 1.91]        & [0.00319, 6.05]  & $1.20 \pm 0.29$   \\  {[7,\textbf{88};17186,52443,7629,11701]}     & 2.40  & [2.06, 2.71]   & [1.39, 3.32]&      n/a	 & n/a & n/a   & $0.769 \pm 0.156$ \\ 
          {[88;7629,11701]}  & 0.822  & [0.143, 1.87]   & [0.00148, 5.92]&       n/a	 & n/a  & n/a   & $0.769 \pm 0.156$ \\ 
          {[89;38057,54846]}   & 1.03 & [0.333, 1.97] & [0.00172, 4.52]  & n/a & n/a  & n/a  & $0.337  \pm 0.092$  \\ 
          {[216;23747,170964]}   & 0.0283 & [0.000112, 0.0927] & [0.000112, 0.417]  & n/a & n/a  & n/a  & $0.233  \pm 0.10$  \\ 
          {[704;43993,48500]}   & 1.40 & [0.651, 2.23] & [0.00149, 4.95]  & 0.155 & [0.00910, 0.664]  & [0.00182, 3.50]  & $1.65  \pm 0.23$  \\ 
    \end{tabular}
  \end{center}
\end{sidewaystable}

Our ML masses are typically close to the average literature values summarized by \citet{Car12}. In most test cases these average values are well within our $1\sigma$ boundaries whereas in \citet{Sil17} the literature values were typically only within the $3\sigma$ boundaries (Tables \ref{main_results_bc430} and \ref{main_results_nobc430}). One interesting exception to these results is (10) Hygiea for which the significant differences between the results with and without BC430 are likely caused by at least one of the asteroids having close encounters with Ceres, Pallas and/or Vesta which, when not considered, lead to inaccurate orbits and, further, masses.

The mass estimates for asteroids (41) Daphne, (89) Julia and (52) Europa have particularly wide uncertainties. The reason is that the perturbations on the test asteroid orbits from the close encounters used are quite weak. \citet{Zie11} previously studied the first two of these targets using a much larger number of test asteroids and obtained significantly lower uncertainties in comparison to ours. We believe that including many more test asteroids would similarly reduce the formal uncertainties of our results for all three asteroids. However, the important aspect for this study is that the latest literature values are within the $1-\sigma$ limits of our results showing that the uncertainty estimates that we obtain are reliable.

\begin{figure}
  \begin{center}
    \includegraphics[width=1.0\columnwidth]{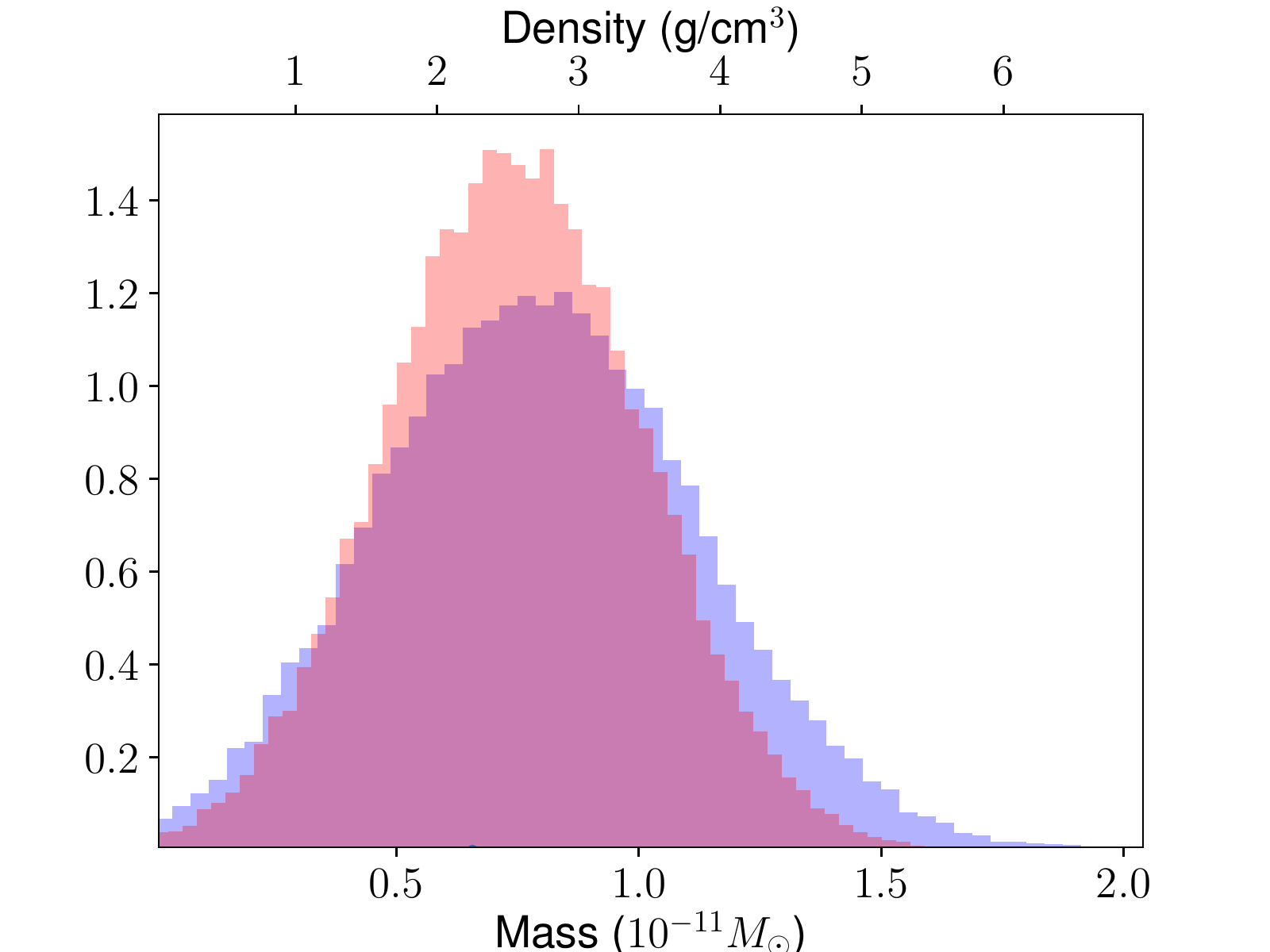}
    \caption{The probability densities for the mass of (16) Psyche with BC430 perturbations for the cases of [16;17799,20837] (blue) and [16;91495,151878] (red). The upper x-axis shows the bulk density corresponding to the equivalent mass on the lower x-axis assuming a volume-equivalent diameter of 223 km  (Table~\ref{main_results_bc430})  corresponding to a volume of $5.806 \times 10^6$ km$^3$.
    We note that the volume uncertainty is not considered in this figure.}
    \label{16_distros}
  \end{center}
\end{figure}

\begin{figure}
  \begin{center} 
    \includegraphics[width=1.0\columnwidth]{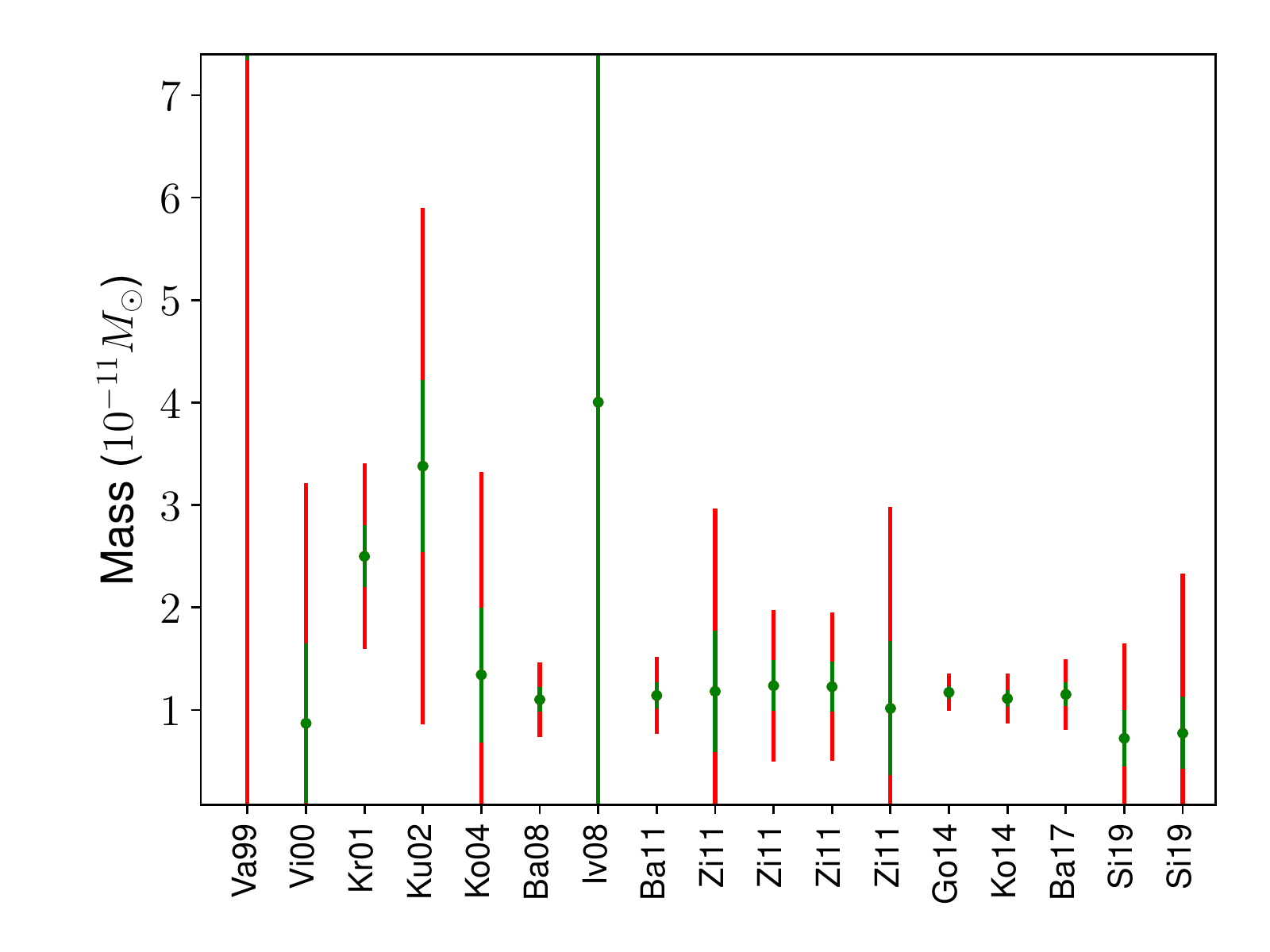}
    \caption{A comparison of previous mass estimates for (16) Psyche based on asteroid-asteroid perturbations (References from left to right: \citet{Vas99,Via00,Kra01,Kuz02,Koc04,Bae08,Iva08,Bae11,Zie11,Gof14,Koc14,Bae17}; this work (Table~\ref{main_results_bc430})). The green error bars represent $1\sigma$ limits while the red error bars represent $3\sigma$ limits. }
    \label{psyche_comparison}
  \end{center}
\end{figure}

Asteroid (16) Psyche is the target of NASA's Psyche mission \citep{Elk14} and is commonly believed to be the iron core of a protoplanet. A recent bulk density estimate of approximately 4~g/cm$^3$ given a mass estimate of $(1.21 \pm 0.16) \times 10^{-11} \msun$ \citep{Vii18} supports the theory that (16) Psyche is an iron core. However, our ML mass is significantly lower than the  above estimate and consequently points to a significantly lower bulk density than that reported in the literature (Fig.~\ref{16_distros}). This literature value nevertheless remains within our $3\sigma$ limits and therefore our result does not rule out a significantly higher bulk density. Due to these interesting results, we chose to do a separate run for (16) Psyche in order to see whether this result can be reproduced with different test asteroids. Our ML masses from the two independent runs are similar which shows that the small mass can indeed be reproduced with a different set of test asteroids (Table~\ref{main_results_bc430} and Fig.~\ref{16_distros}). We have also included a comparison between our results and previous studies \citep{Vii18} for Psyche (Fig.~\ref{psyche_comparison}) which confirms that our nominal masses are indeed lower than all previous literature values whereas our uncertainties are not unusually wide for this object.

\begin{figure}
  \begin{center} 
    \includegraphics[width=1.0\columnwidth]{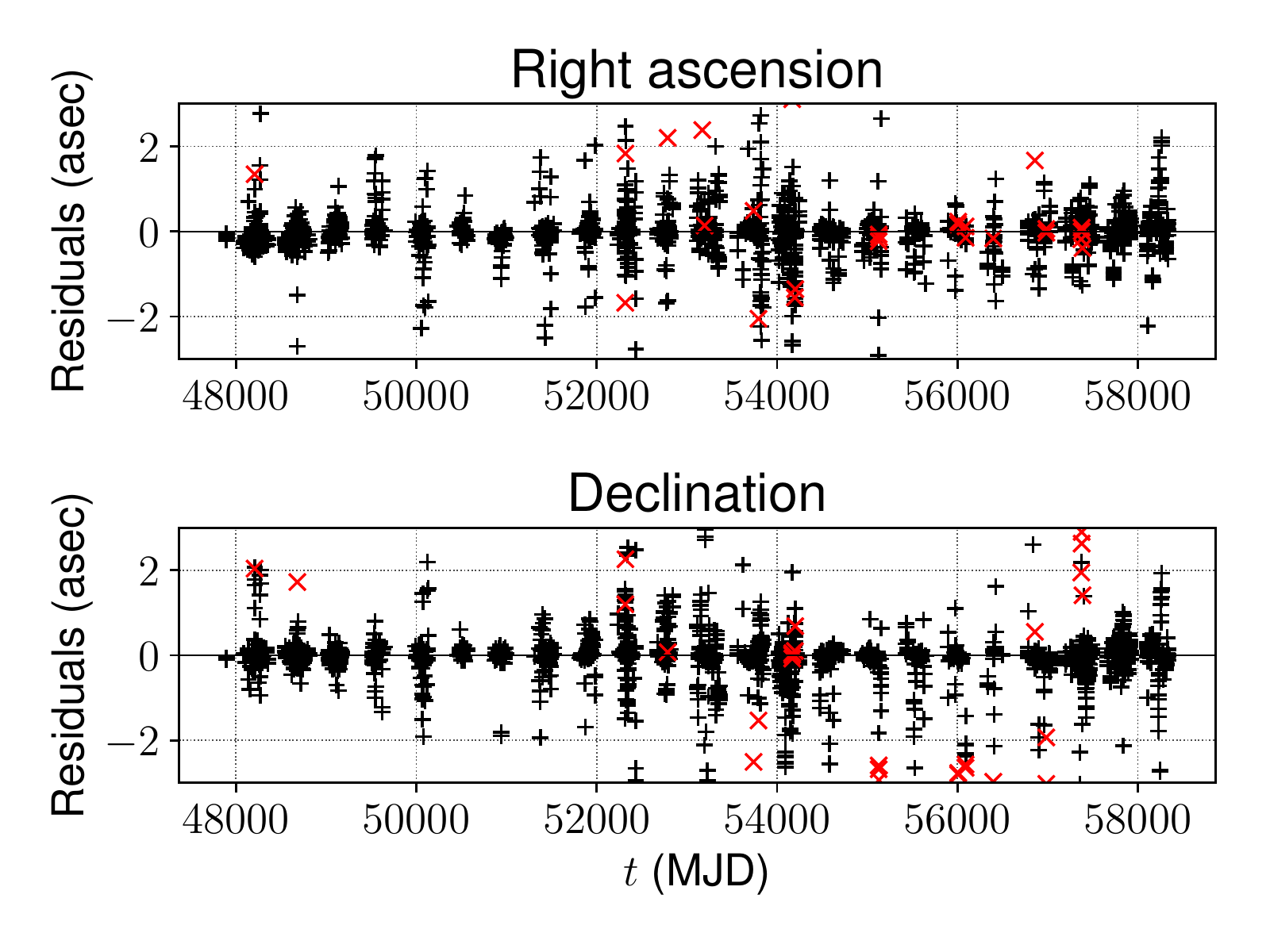}
    \caption{Mean residuals of all accepted MCMC proposals for asteroid (16) Psyche in the case of [16;17799,20837].
    The red crosses represent data that has been rejected as outliers and some outliers are entirely outside of the plot boundaries.} 
    \label{16_resids}
  \end{center}
\end{figure}

\begin{figure}
  \begin{center} 
    \includegraphics[width=1.0\columnwidth]{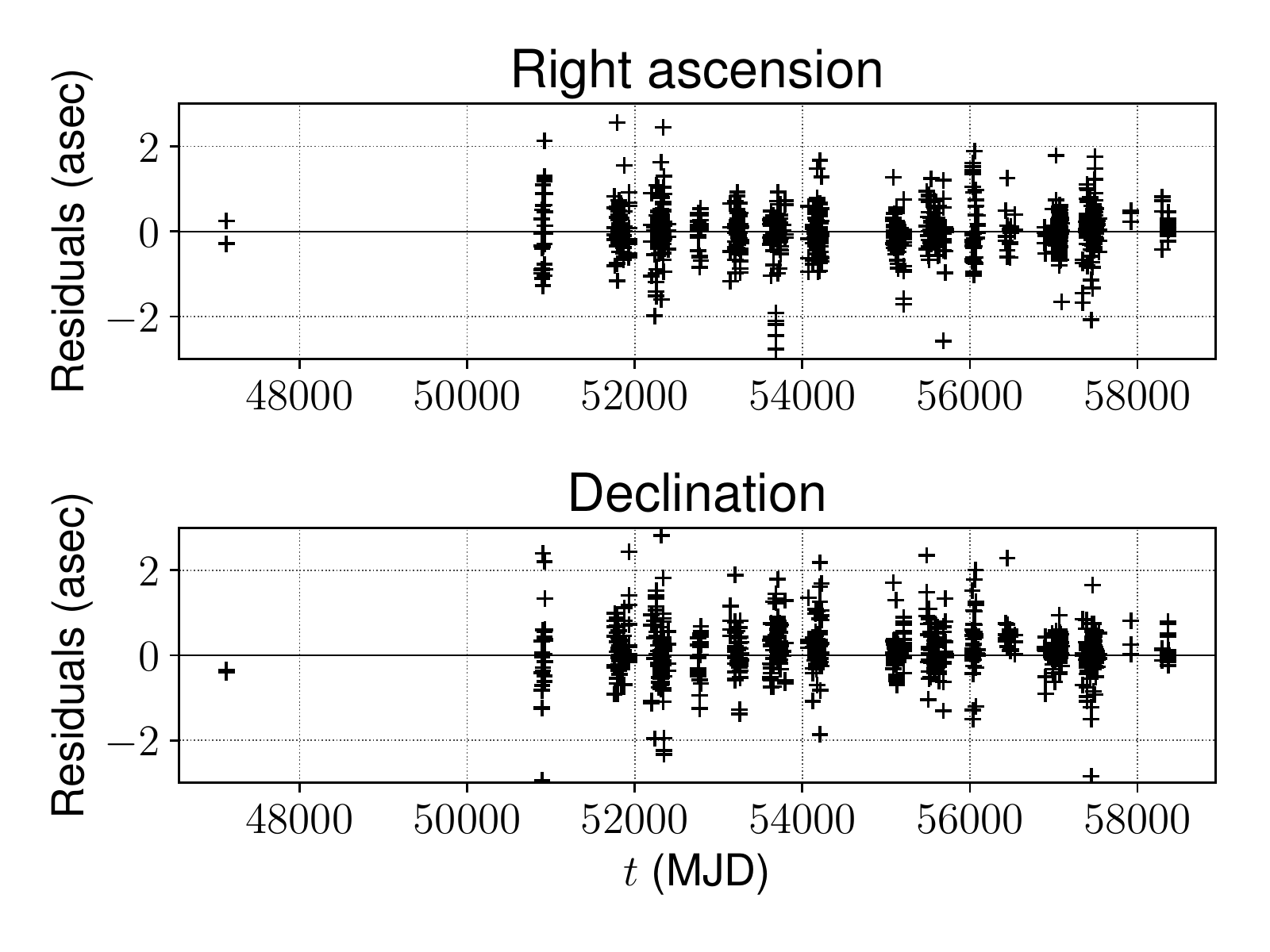}
    \caption{Mean residuals of all accepted MCMC proposals for asteroid (17799) Petewilliams in the case of [16;17799,20837].} 
    \label{17799_resids}
  \end{center}
\end{figure}

\begin{figure}
  \begin{center} 
    \includegraphics[width=1.0\columnwidth]{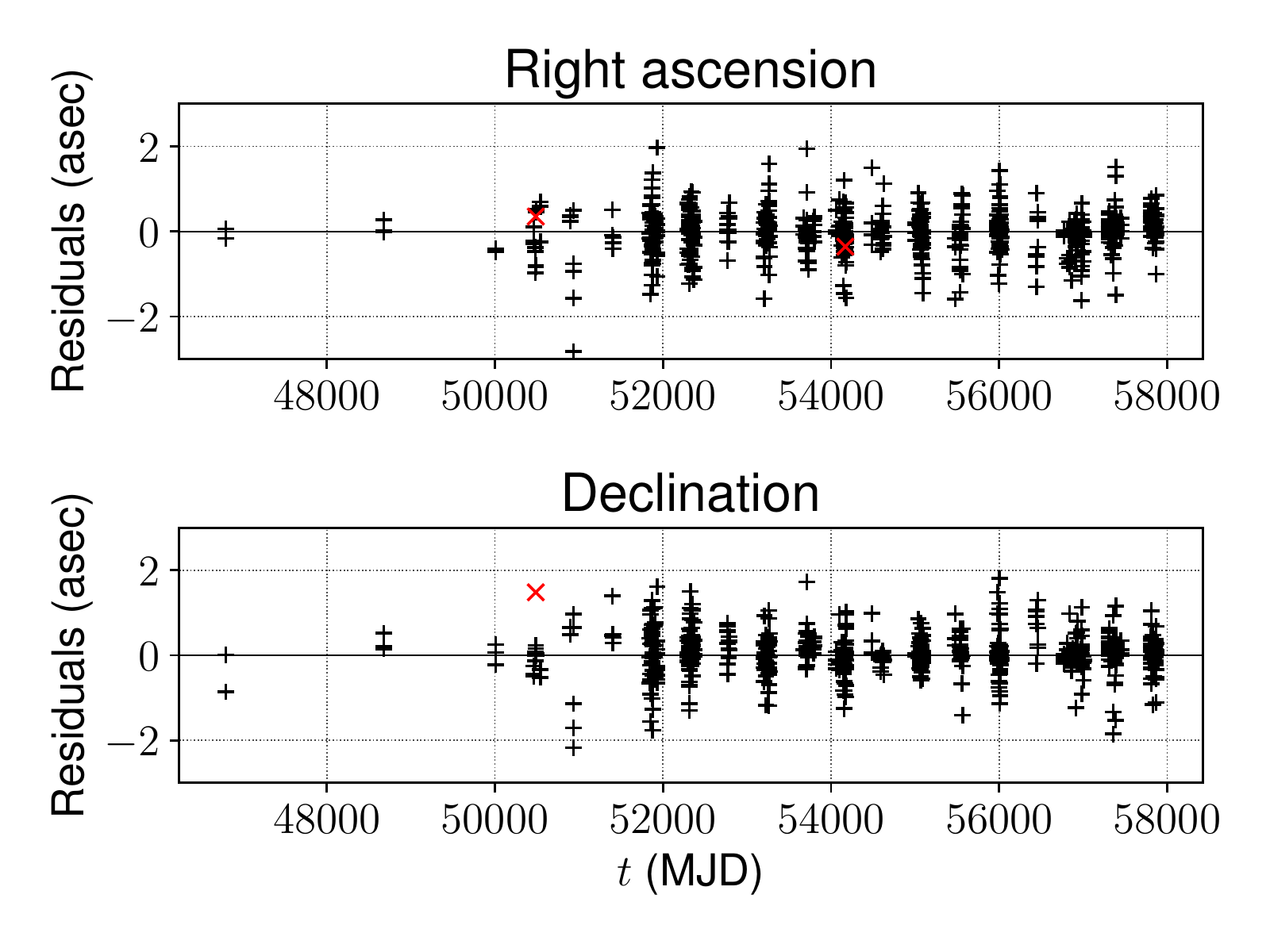}
    \caption{Mean residuals of all accepted MCMC proposals for asteroid (20837) Ramanlal  in the case of [16;17799,20837].
    The red crosses represent data that has been rejected as outliers and some outliers are entirely outside of the plot boundaries.}
    \label{20837_resids}
  \end{center}
\end{figure}

The mean residuals for each asteroid in the [16;17799,20837] case show that there are no obvious systematic effects visible in the residuals that could explain the small ML mass (Figs.~\ref{16_resids}, \ref{17799_resids} and \ref{20837_resids}). We note that a few residuals are beyond the scale of the plots.

\begin{figure}
\begin{center}
    \includegraphics[width=1.0\columnwidth]{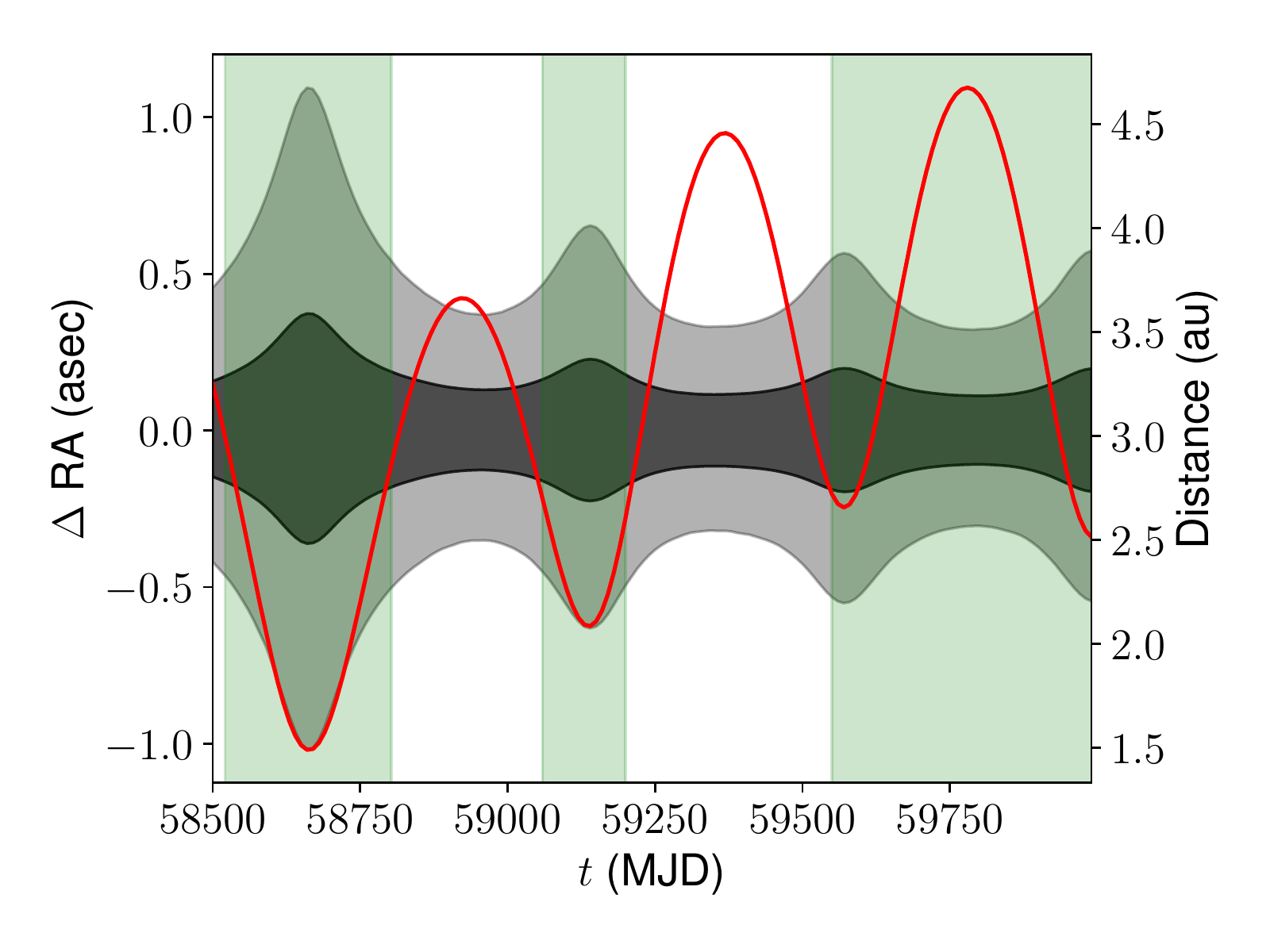}
    \caption{Ephemeris prediction for asteroid (151878) 2003 PZ$_4$ up to MJD 60000 in terms of RA relative to the best-fit value. The $1\sigma$ and $3\sigma$ credible intervals are shown in darker and lighter gray, respectively. The red line shows the asteroid's topocentric distance as a function of time. The green color represents times when the asteroid is observable assuming a topocentric observer by requiring that the solar elongation is greater than 60 degrees and the apparent V magnitude is less than 21.}
    \label{151878_ra_prediction}
  \end{center}
\end{figure}

\begin{figure}
\begin{center}
    \includegraphics[width=1.0\columnwidth]{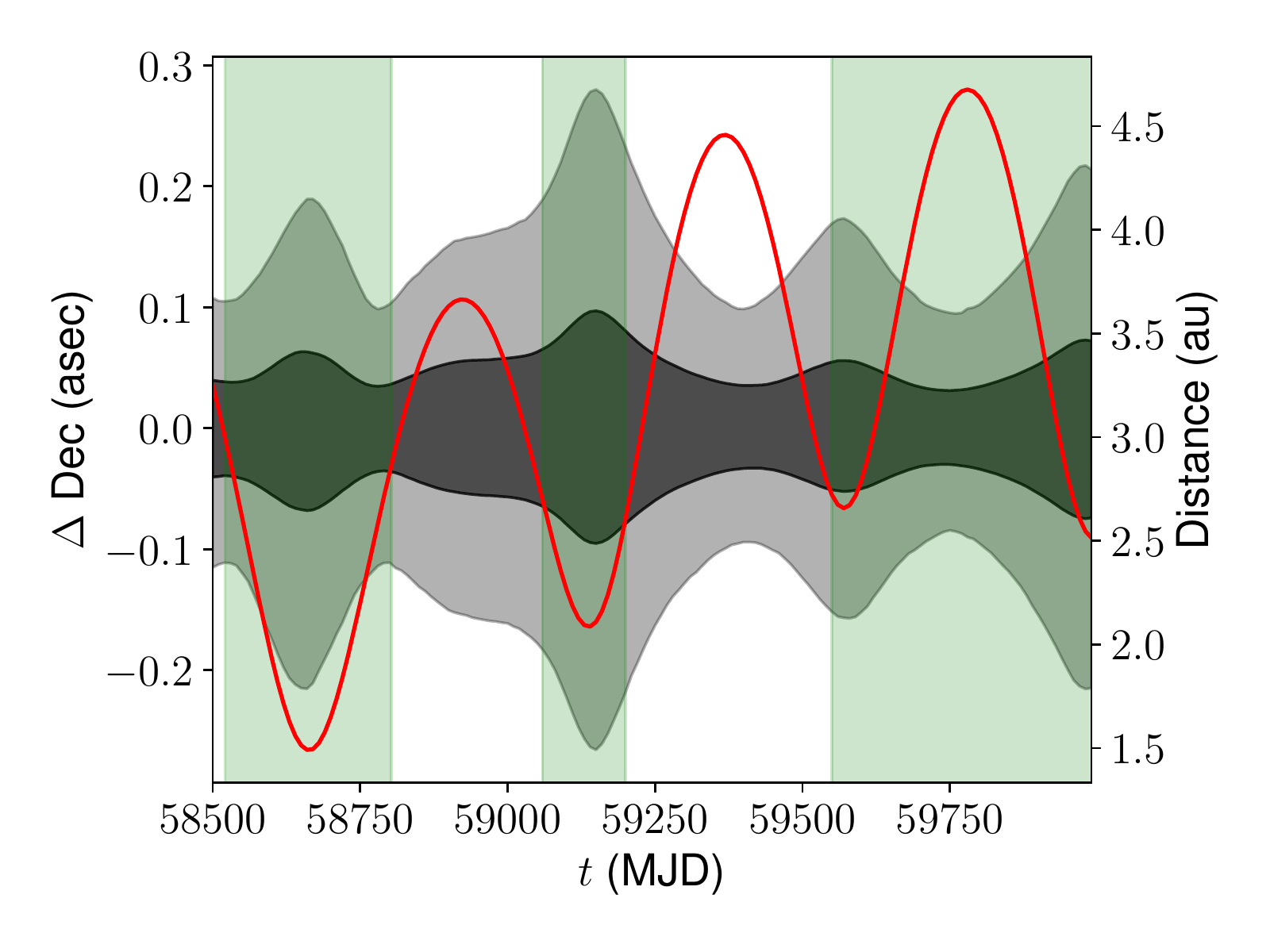}
   \caption{As Fig.~\ref{151878_ra_prediction} but for Dec.}
    \label{151878_dec_prediction}
  \end{center}
\end{figure}

\begin{figure}
\begin{center}
    \includegraphics[width=1.0\columnwidth]{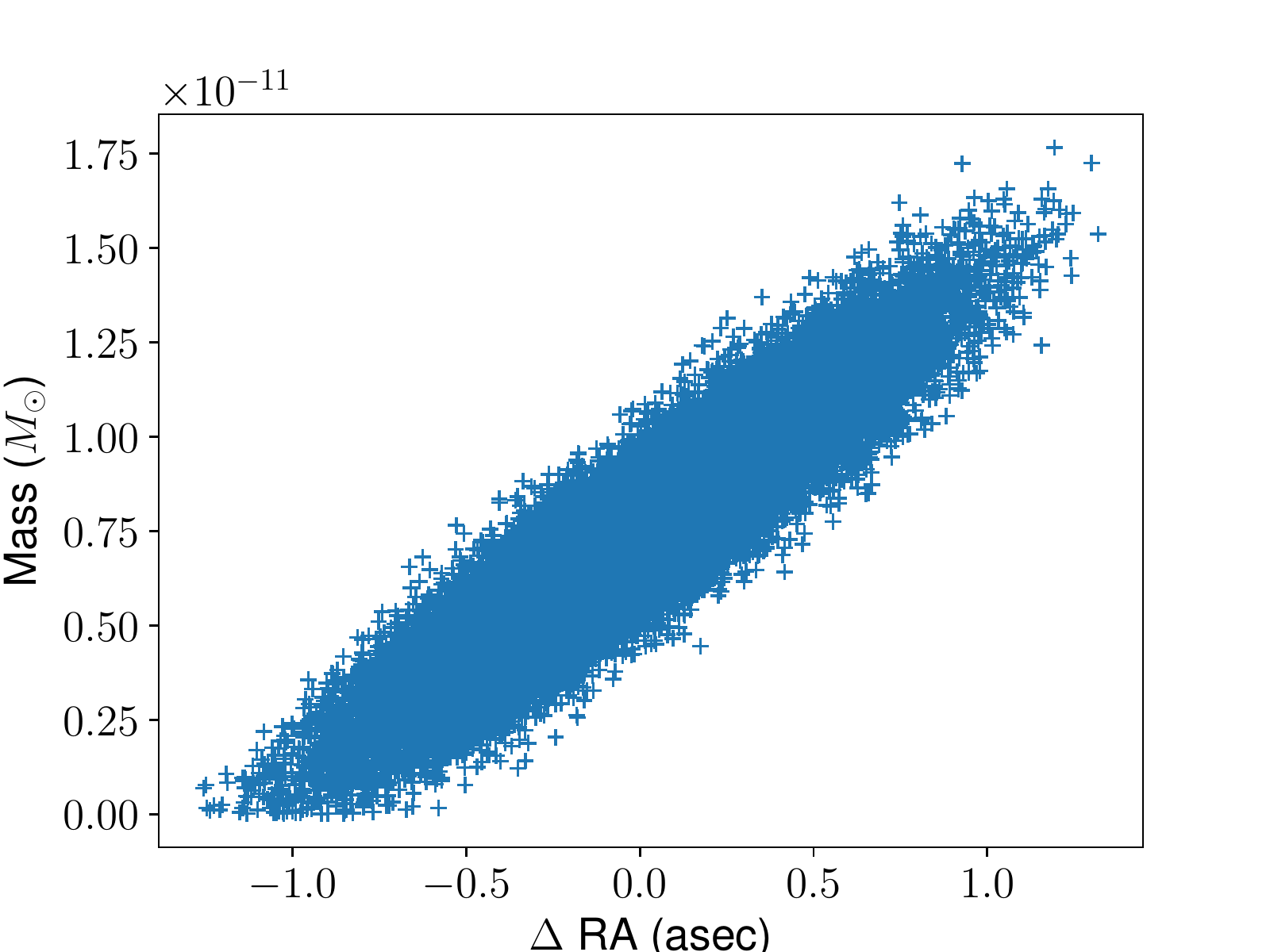}
    \caption{Mass of (16) Psyche versus the RA prediction for asteroid (151878) 2003 PZ$_4$ relative to the best-fit value at MJD 58650.}
    \label{151878_ra_58650}
  \end{center}
\end{figure}

\begin{figure}
\begin{center}
    \includegraphics[width=1.0\columnwidth]{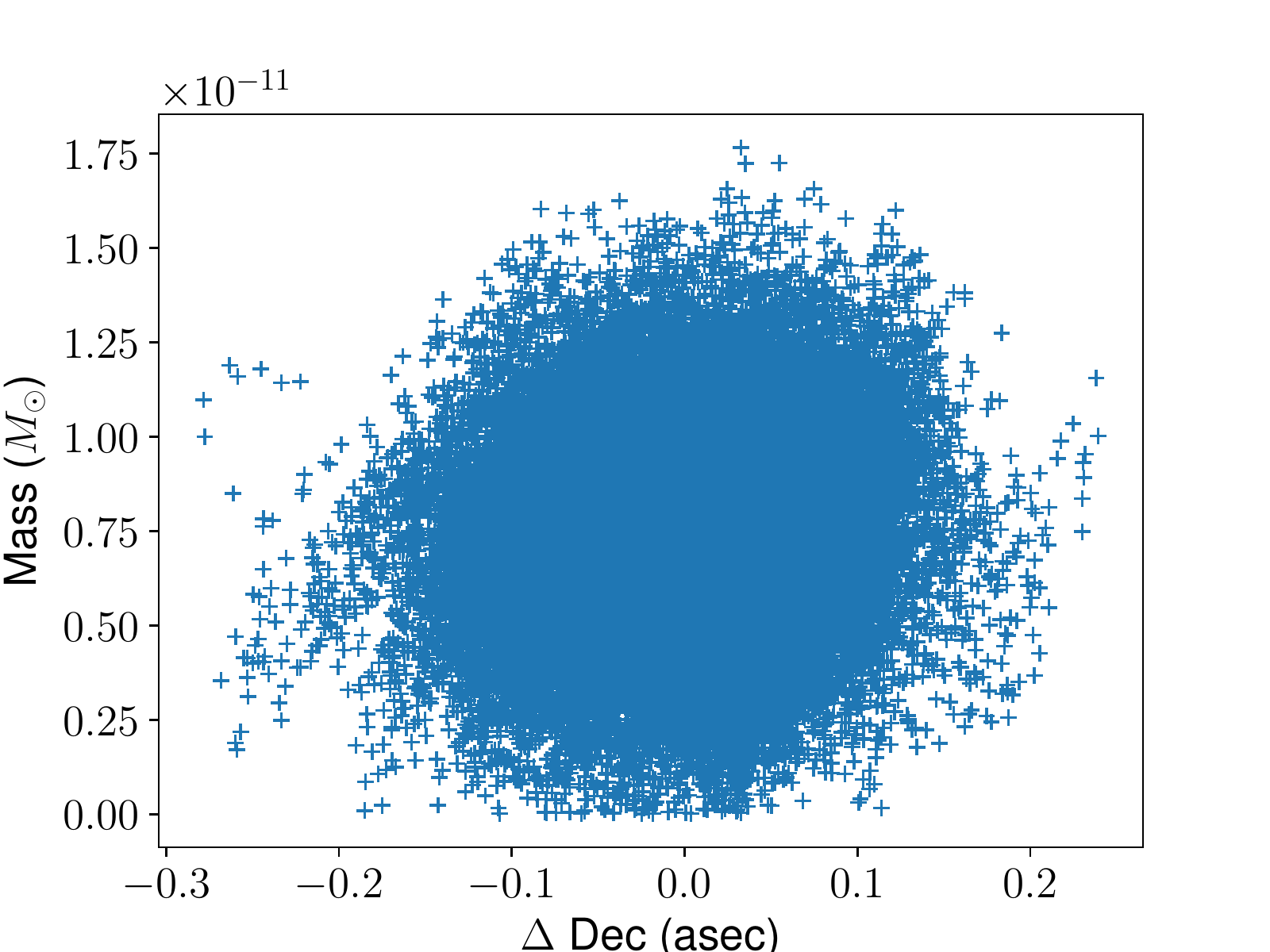}
    \caption{As Fig.~\ref{151878_ra_58650} but for Dec instead of RA.}
    \label{151878_dec_58650}
  \end{center}
\end{figure}

To investigate the case of (16) Psyche further we computed topocentric ephemerides for test asteroid (151878) 2003 PZ$_4$ in the case of [16;91495,151878] several years into the future and include the perturbations caused by (16) Psyche. This lets us quantify the impact of (16) Psyche's perturbations on the test asteroid's orbit and to assess when astrometry should be obtained to efficiently constrain our mass estimate for (16) Psyche (Figs.~\ref{151878_ra_prediction} and \ref{151878_dec_prediction}). The behavior of RA around MJD 58650, corresponding to June 2019, where different perturber masses lead to ephemerides with differences up to two arcseconds due to the asteroid's small distance from Earth, is particularly interesting. The uncertainty in the ephemeris prediction is driven by the uncertainty in perturber mass: a larger mass corresponds to a larger right ascension, while for declination there is little correlation (Figs.~\ref{151878_ra_58650} and \ref{151878_dec_58650}). In practice, this means that high-accuracy astrometry of asteroid (151878) 2003 PZ$_4$ obtained during the summer of 2019 will significantly reduce the uncertainty of the mass estimate. The asteroid is easily observable during that time and we thus expect that suitable astrometry will be obtained by asteroid surveys. 

Although our ML masses have improved, partly due to additional data, our uncertainties tend to be wider than in \citet{Sil17} (Table~\ref{main_results_bc430}). 
We believe this is largely caused by the different observational weights used; in particular, the $\sqrt{N}$ term, which was not used in our previous work, directly leads to much looser weights for some observations which translates to greater uncertainties in the results. To demonstrate this effect we also chose to test MCMC without the $\sqrt{N}$ factor applied in the error model (Table~\ref{nosqrt_results}). Upon comparison to results with the factor included (Table~\ref{main_results_bc430}) it is clear that the factor does indeed increase the uncertainties  of our test cases as expected.

To assess the impact of multiple simultaneous test asteroids on our results, we also chose to run our algorithm for each test asteroid separately for some of these cases (Tables \ref{onetest_bc430} and \ref{onetest_nobc430}). In general, these results also show clear improvement in comparison to our previous study, as the ML masses now tend to be significantly closer to the average literature value. We do, however, note the case of [10;1259] where our ML mass is clearly unrealistic in Table~\ref{onetest_nobc430}, being an order of magnitude above even the mass of Ceres. [29;7060] and [52;306] exhibit similar behavior, but these two cases also have extremely wide uncertainties and the previous literature values easily fall within these. We suspect that in the latter two cases the perturbations simply are very weak, directly resulting in very high noise, making these particular encounters quite poor for mass estimation. Indeed, comparing these values to those in Table~\ref{main_results_nobc430} we see that the 'better' encounters dominate in the cases with both test asteroids combined, which is to be expected. 

The inclusion of perturbations of Ceres, Pallas and Vesta has a significant impact on [10;1259] (Table~\ref{onetest_bc430}); where 
previously the ML mass was an order of magnitude above that of Ceres, it is now much closer to the literature values, though the noise level remains high. This appears
to confirm that (1259) Ogyalla has a close encounter with at least one of these three perturbers during the observational timespan, though we are not aware of any previously published information on such an encounter. The cases of [29;7060] and [52;306], however, remain extremely noisy.

\begin{figure}
 \begin{center} 
    \includegraphics[width=1.0\columnwidth]{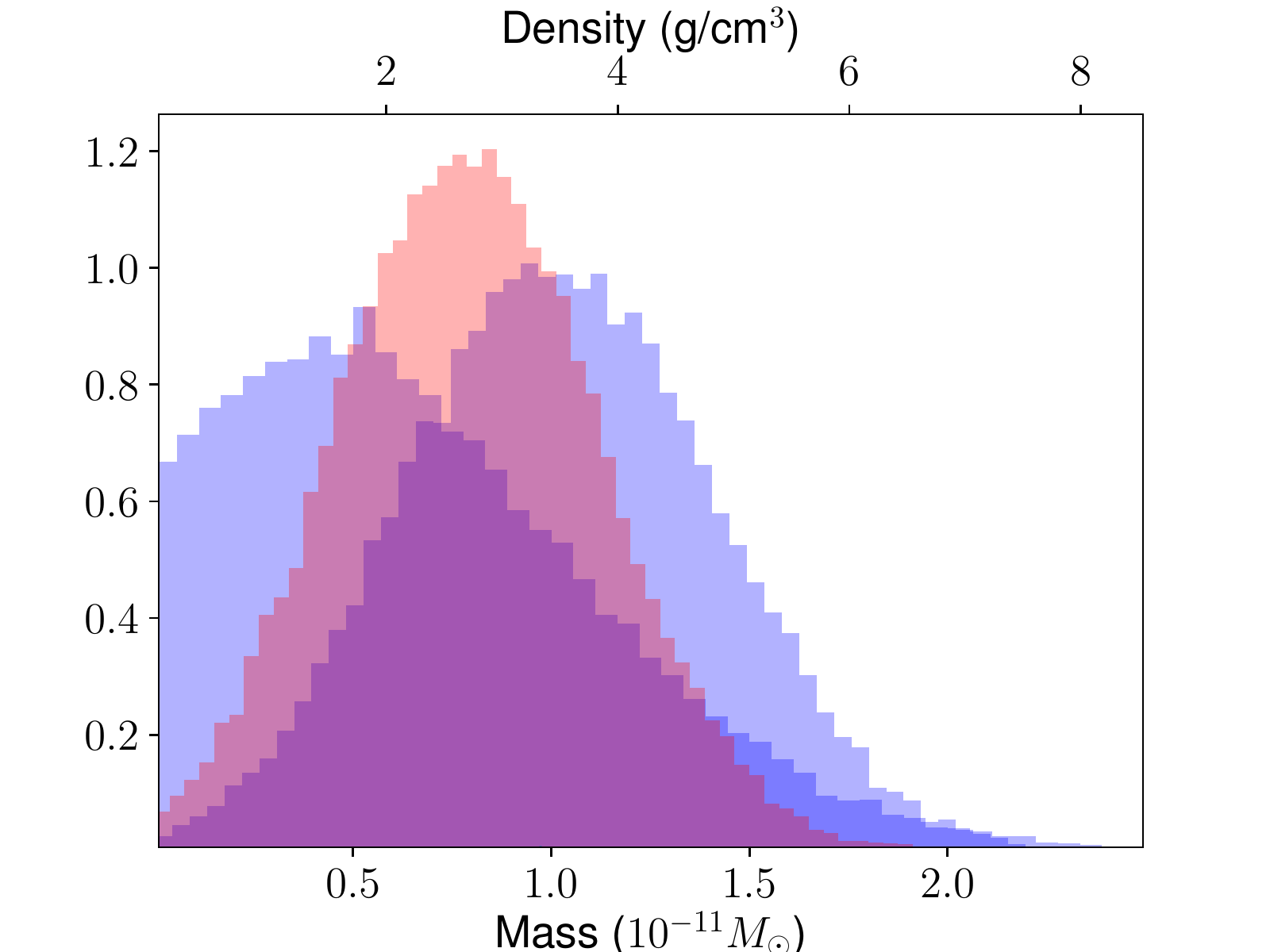}
    \caption{The probability densities for the mass of (16) Psyche with BC430 perturbations for two simultaneous test asteroids (17799 and 20837, red) and two cases with each test asteroid considered separately (blue). The upper x-axis shows the bulk density corresponding to the equivalent mass on the lower x-axis assuming a volume-equivalent diameter of 223 km \citep{Dru18}}
    \label{16_single_vs_multi}
 \end{center}
\end{figure}

To show how two simultaneous test asteroids improves the mass estimate compared to two separate runs with a single test asteroid each, for the example case of (16) Psyche we have included a figure showcasing our results with two simultaneous test asteroids compared with results for each test asteroid considered separately  (Fig.~\ref{16_single_vs_multi}). The peak of the distribution for both test asteroids combined corresponds to the peak of the overlapping part of the separate single-test-asteroid distributions, which was our expectation. This essentially means that the maximum likelihood value for the mass in the former case corresponds to the mass that best fits both separate test asteroids simultaneously. 

In order to study how increasing the number of data points affects the resulting mass distributions, we opted to perform MCMC runs with approximately half the observational timespan to limit the number of astrometry used for the cases of [15;765] and [15;765,14401] (Table~\ref{eunomia_test}) where, for reference, the encounter with the former test asteroid took place in 2010 and the latter in 2005 \citep{Gal02}. Here one can see that for the case of [15;765,14401] cutting the observational timespan to half slightly increases the uncertainty of the mass, while for [15;765] alone the change is slightly greater. Using a timespan of 2003-2007 the result for [15;765] becomes essentially useless, which is expected given that no close encounters between these two asteroids exist in this time period. While we did not perform the computation for [15;14401] in this case, it is clear that this test asteroid dominates the results of [15;765,14401] and would be expected to provide essentially the same result. 
Overall, these results show that indeed additional data does significantly reduce the uncertainties of the mass, as is expected.

\begin{table}
\footnotesize
  \caption{Results of the MCMC algorithm for data without the $\sqrt{N}$ factor. The perturbations of Ceres, Pallas, and Vesta are included.}
  \label{nosqrt_results} 
  \begin{center}
  \renewcommand{\arraystretch}{1.2}
    \begin{tabular}{cccccccccc} 
      Encounter	   &  ML mass              & $1\sigma$ boundaries   & $3\sigma$ boundaries  \\
      & [$10^{-11} \msun$] & [$10^{-11} \msun$]   & [$10^{-11} \msun$]   \\ \hline 
     {[15;765,14401]} & 0.733  & [0.619, 0.854]  & [0.387, 1.07] \\
     {[16;17799,20837]}  & 0.726  & [0.520, 0.925]   & [0.124, 1.34] \\
     {[16;91495,151878]}  & 0.711 & [0.552, 0.874]   & [0.235, 1.19] \\


    \end{tabular}
  \end{center}
\end{table}
\begin{table*}
\footnotesize
  \caption{Compilation of the MCMC algorithm's results including perturbations of Ceres, Pallas and Vesta for several cases from our real data set separately for each test asteroid.
    The first reference masses and their uncertainties are taken from our previous work \citep{Sil17} whereas the second are weighted averages of previous literature values \citep{Car12}}
  \label{onetest_bc430} 
  \begin{center}
    \begin{tabular}{cccccccc}
      \hline
      Encounter	  & ML mass              & $1\sigma$ boundaries   & $3\sigma$ boundaries  & Ref. mass~1  & $1\sigma$ boundaries   & $3\sigma$ boundaries & Ref. mass~2  \\
      & [$10^{-11} \msun$] & [$10^{-11} \msun$]   & [$10^{-11} \msun$]   & [$10^{-11} \msun$] & [$10^{-11} \msun$] & [$10^{-11} \msun$] & [$10^{-11} \msun$] \\ \hline 
          {[10;1259]}   & 10.3  & [3.57, 25.1]   & [0.0200, 66.1]    & 2.48 & [2.21, 2.77] & [1.63, 3.32]    & $4.34 \pm 0.26$ \\ 
          {[10;57493]}   & 5.94  & [4.20, 7.70]   & [0.795, 11.1]    & 2.48 & [2.21, 2.77] & [1.63, 3.32]    & $4.34 \pm 0.26$ \\ 
          {[15;14401]}   & 1.55  & [1.29, 1.81]   & [0.777, 2.33]  &     1.11	 & [0.914, 1.25]  & [0.574, 1.61]   & $1.58 \pm 0.09$ \\ 
          {[15;765]}     & 1.57  & [1.14, 2.01]   & [0.261, 2.87]&      1.11	 & [0.914, 1.25]  & [0.574, 1.61]   & $1.58 \pm 0.09$ \\ 
          {[16;17799]}    & 0.463  & [0.0758, 1.01]   & [0.000872, 3.33]&       n/a	 & n/a  & n/a   & $1.37 \pm 0.38$ \\ 
          {[16;20837]}    & 1.01  & [0.602, 1.42]   & [0.00647, 2.64]&       n/a	 & n/a  & n/a   & $1.37 \pm 0.38$ \\ 
          {[19;3486]}   & 0.429 & [0.206, 0.667] & [0.000359, 1.41] & 0.141 & [0.0567, 0.285]  & [0.000467, 0.828] & $0.433 \pm 0.073$   \\ 
          {[19;27799]}   & 0.362 & [0.0841, 0.719] & [0.0005334, 2.16] & 0.141 & [0.0567, 0.285]  & [0.000467, 0.828] & $0.433 \pm 0.073$   \\ 
          {[29;987]}     & 0.547 & [0.0151, 1.47] & [0.00129, 5.11]  & 0.258	& [0.0163, 0.898]  & [0.00238, 4.43] & $0.649 \pm 0.101$  \\ 
          {[29;7060]}     & 23.2 & [1.51, 59.2] & [0.0560, 233]  & 0.258	& [0.0163, 0.898]  & [0.00238, 4.43] & $0.649 \pm 0.101$  \\ 
          {[52;124]}      & 1.30  & [0.0381, 3.36]  & [0.00298, 10.22] & 0.893 & [0.232, 1.91]        & [0.00319, 6.05]  & $1.20 \pm 0.29$   \\ 
          {[52;306]}      & 25.2  & [5.53, 54.6]  & [0.0403, 159] & 0.893 & [0.232, 1.91]        & [0.00319, 6.05]  & $1.20 \pm 0.29$   \\ 
    \end{tabular}
  \end{center}
\end{table*}

\begin{table*}
\footnotesize
  \caption{Compilation of the MCMC algorithm's results without the perturbations of Ceres, Pallas, and Vesta for several cases from our real data set separately for each test asteroid.
    The first reference masses and their uncertainties are from \citep{Sil17} whereas the second are weighted averages of literature values \citep{Car12}}
  \label{onetest_nobc430} 
  \begin{center}
    \begin{tabular}{cccccccc}
      \hline
      Encounter	  & ML mass              & $1\sigma$ boundaries   & $3\sigma$ boundaries  & Ref. mass~1  & $1\sigma$ boundaries   & $3\sigma$ boundaries & Ref. mass~2  \\
      & [$10^{-11} \msun$] & [$10^{-11} \msun$]   & [$10^{-11} \msun$]   & [$10^{-11} \msun$] & [$10^{-11} \msun$] & [$10^{-11} \msun$] & [$10^{-11} \msun$] \\ \hline 
          {[10;1259]}   & 108  & [92.8, 123]   & [63.3, 153]    & 2.48 & [2.21, 2.77] & [1.63, 3.32]    & $4.34 \pm 0.26$ \\ 
          {[10;57493]}   & 5.53  & [3.83, 7.26]   & [0.567, 10.6]    & 2.48 & [2.21, 2.77] & [1.63, 3.32]    & $4.34 \pm 0.26$ \\
          {[16;17799]}    & 0.467  & [0.0791, 1.08]   & [0.000922, 3.60]&       n/a	 & n/a  & n/a   & $1.37 \pm 0.38$ \\ 
          {[16;20837]}    & 1.09  & [0.675, 1.48]   & [0.00241, 2.30]&       n/a	 & n/a  & n/a   & $1.37 \pm 0.38$ \\ 
          {[19;3486]}   & 0.284 & [0.0973, 0.492] & [0.000319, 1.26] & 0.141 & [0.0567, 0.285]  & [0.000467, 0.828] & $0.433 \pm 0.073$   \\ 
          {[19;27799]}   & 0.337 & [0.0779, 0.694] & [0.000551, 2.08] & 0.141 & [0.0567, 0.285]  & [0.000467, 0.828] & $0.433 \pm 0.073$   \\ 
          {[29;987]}     & 0.540 & [0.0218, 1.40] & [0.00136, 5.33]  & 0.258	& [0.0163, 0.898]  & [0.00238, 4.43] & $0.649 \pm 0.101$  \\ 
          {[29;7060]}     & 30.3 & [4.37, 71.0] & [0.0602, 254]  & 0.258	& [0.0163, 0.898]  & [0.00238, 4.43] & $0.649 \pm 0.101$  \\ 
          {[52;124]}      & 1.36  & [0.0739, 3.44]  & [0.00316, 10.3] & 0.893 & [0.232, 1.91]        & [0.00319, 6.05]  & $1.20 \pm 0.29$   \\ 
          {[52;306]}      & 29.6  & [7.47, 60.8]  & [0.0429, 163] & 0.893 & [0.232, 1.91]        & [0.00319, 6.05]  & $1.20 \pm 0.29$   \\ 

    \end{tabular}
  \end{center}
\end{table*}

\begin{table*}
\footnotesize
  \caption{Comparison of MCMC results of [15;765,14401] and |15;765] with differing observational timespans. The perturbations of Ceres, Palla and Vesta are included.}
  \label{eunomia_test} 
  \begin{center}
  \renewcommand{\arraystretch}{1.2}
    \begin{tabular}{cccccccccc} 
      Encounter	   & Timespan & ML mass              & $1\sigma$ boundaries   & $3\sigma$ boundaries  \\
      &  &  [$10^{-11} \msun$] & [$10^{-11} \msun$]   & [$10^{-11} \msun$]   \\ \hline 
     {[15;765,14401]}  & 1990-2019 & 1.40  & [1.20, 1.62]   & [0.792, 2.04] \\
     {[15;765,14401]}  & 2000-2015 & 1.20  & [0.879, 1.53]  & [0.262, 2.19] \\
     {[15;765,14401]}  & 2003-2007 & 3.11  & [0.891, 6.05]  & [0.00839, 16.1] \\
     {|15;765]}        & 1990-2019 & 1.57  & [1.15, 2.01]  & [0.260, 2.86] \\
     {|15;765]}        & 2000-2015 & 1.59  & [1.01, 2.16]   & [0.00656, 3.335]   \\
     {|15;765]}        & 2003-2007 & 764  & [114, 1759]   & [1.69, 6397]   \\
    \end{tabular}
  \end{center}
\end{table*}

\section{Conclusions}

We have successfully developed, implemented, and validated a robust adaptive Metropolis algorithm for asteroid mass estimation. Our results show significant improvement compared to our previous work \citep{Sil17} with the new RAM algorithm, the ability to include multiple perturbers and/or test asteroids simultaneously, the improved force model, and the new observational weighting scheme. The uncertainties remain wider than recent previous literature values while in almost all cases the literature values are within our $1\sigma$ limits.

The interesting exception is (16) Psyche for which we obtain a significantly smaller mass than previously reported in the literature. Given that we find consistent results with two independent data sets---less the data set for (16) Psyche itself---this result merits further analysis when additional data on (151878) 2003 PZ$_4$ is obtained later this year.

In the future, we intend to extend OpenOrb so that we can use Gaia astrometry in combination with ground-based astrometry for asteroid mass estimation.

\section*{Acknowledgements}
This work was supported by grants \#299543, \#307157, and \#328654 from the Academy of Finland. This research has made use of NASA's Astrophysics Data System, and data and/or services provided by the International Astronomical Union's Minor Planet Center. We thank the anonymous referee for their useful comments which greatly improved our paper. We would also like to thank James Baer and William Zielenbach for their answers to our questions about their respective works and the latter also for providing us with lists of close encounters with three of our target perturbers. The computations for this work were mainly performed with the Kale cluster of the Finnish Grid and Cloud infrastructure  (persistent identifier: urn:nbn:fi:research-infras-2016072533), and the scientific computing environment of the Nordic Optical Telescope set up by Peter Sorensen.

\bibliographystyle{apa}
\bibliography{refs.bib}

\end{document}